\documentclass[preprint]{aastex}
\usepackage{epsf}
\newcommand{\simgt}{\lower.5ex\hbox{$\; \buildrel > \over \sim \;$}}
\newcommand{\simlt}{\lower.5ex\hbox{$\; \buildrel < \over \sim \;$}}
\newcommand{\himpc}{{\hbox {$h^{-1}$}{\rm Mpc}} }
\newcommand{\deltah}{{\delta_{\rm halo}}}
\newcommand{\deltam}{{\delta_{\rm mass}}}
\newcommand{\deltag}{{\delta_{\rm gal}}}
\newcommand{\gal}{{\Delta_{\rm g}}}
\newcommand{\halo}{{\Delta_{\rm h}}}
\newcommand{\sigmamm}{{\sigma_{\rm mm}}}

\newcommand{\sigmagg}{{\sigma_{\rm gg}}}
\newcommand{\bhat}{{\hat{b}}}
\newcommand{\btilde}{{\tilde{b}}}
\newcommand{\sigmab}{{\sigma_{\rm b}}}
\newcommand{\bvar}{{b_{\rm var}}}
\newcommand{\rcorr}{{r_{\rm corr}}}
\newcommand{\scatt}{{\epsilon_{\rm scatt}}}
\newcommand{\nonl}{{\epsilon_{\rm nl}}}
\newcommand{\bcov}{{b_{\rm cov}}}
\newcommand{\zf}{{z_{\rm f}}}
\newcommand{\sigmahm}{{\sigma_{\rm hm}}}
\newcommand{\sigmahh}{{\sigma_{\rm hh}}}
\newcommand{\msun}{{M_{\odot}}}
\newcommand{\llangle}{{\langle\hspace*{-1.0mm}\langle}}
\newcommand{\rrangle}{{\rangle\hspace*{-1.0mm}\rangle}}
\newcommand{\MW}{ {\scriptscriptstyle {\rm MW}} }
\slugcomment{RESCEU-5/2000\quad UTAP-362/2000}
\shorttitle{Nonlinear stochastic biasing}
\shortauthors{Taruya \& Suto}
\begin{document}
%
\title{Nonlinear stochastic biasing \\
from the formation epoch distribution of dark halos
}
%
\author{Atsushi Taruya and Yasushi Suto}
\affil{Department of Physics and Research Center for
    the Early Universe (RESCEU) \\ School of Science, University of
    Tokyo, Tokyo 113-0033, Japan.}
\email{ataruya@utap.phys.s.u-tokyo.ac.jp, suto@phys.s.u-tokyo.ac.jp}
%
\received{2000 February 9}
\accepted{2000 April 19}
\begin{abstract}
  We propose a physical model for nonlinear stochastic biasing of one-point
  statistics resulting from the formation epoch distribution of dark
  halos. In contrast to previous works on the basis of extensive
  numerical simulations, our model provides for the first time an
  analytic expression for the joint probability function.  Specifically
  we derive the joint probability function of halo and mass density
  contrasts from the extended Press-Schechter theory. Since this
  function is derived in the framework of the standard gravitational
  instability theory assuming the random-Gaussianity of the primordial
  density field alone, we expect that the basic features of the
  nonlinear and stochastic biasing predicted from our model are fairly
  generic. As representative examples, we compute the various biasing
  parameters in cold dark matter models as a function of a redshift and
  a smoothing length. Our major findings are (1) the biasing of the
  variance evolves strongly as redshift while its scale-dependence is
  generally weak and a simple linear biasing model provides a reasonable
  approximation roughly at $R\simgt 2(1+z)\himpc$, and (2) the
  stochasticity exhibits moderate scale-dependence especially on
  $R\simlt 20\himpc$, but is almost independent of $z$.  Comparison with
  the previous numerical simulations shows good agreement with the above
  behavior, indicating that the nonlinear and stochastic nature of the
  halo biasing is essentially understood by taking account of the
  distribution of the halo mass and the formation epoch.
\end{abstract} 
%
%
\section{Introduction}
\label{sec:intro}

The universe after the last scattering can be probed only through
observing the distribution of luminous objects, either directly or
indirectly via the weak lensing effect. This is why several wide-field
and/or deep surveys of galaxies, clusters and quasars are planned and
operating at various wavelengths. The purpose of such cosmological
surveys is two-fold; to understand the nature of the astronomical
objects themselves, and to extract the cosmological information. From
the latter point of view, which we will pursue throughout this paper,
the objects serve merely as tracers of dark matter in the universe.
Since such {\it luminous} objects should have been formed as a
consequence of complicated astrophysical processes in addition to the
purely gravitational interaction, it is quite unlikely that they
faithfully trace either the spatial distribution of dark matter or its
redshift evolution.  Rather it is natural to assume that they sample
the dark matter distribution in a biased manner.

To describe the {\it biasing} more specifically, define the density
contrasts of galaxies and mass at a position ${\bf x}$ and redshift
$z$ smoothed over the scale $R$ as
\begin{equation}
\label{eq:dgdm}
\deltag({\bf x},z|R) ={n_{\rm gal}({\bf x},z|R) 
         \over \bar{n}_{\rm gal}} - 1,
\qquad
\deltam({\bf x},z|R) ={\rho_{\rm mass}({\bf x},z|R) 
\over \bar\rho_{\rm mass}} - 1.
\end{equation}
Here and in what follows we use the words, ``mass'' and ``dark
matter'', interchangeably, and ``galaxies'' to imply luminous objects
(galaxies, clusters, quasars, etc.) in a collective sense.

The simplest, albeit most {\it unlikely}, possibility is that they are
proportional to each other:
\begin{equation}
\label{eq:linbias}
\deltag({\bf x},z|R) = b~ \deltam({\bf x},z|R) .
\end{equation}
While the proportional coefficient was assumed to be constant when the
concept of the biasing was first introduced in the cosmology community
(Kaiser 1984; Davis et al. 1985; Bardeen et al.1986), it has been
subsequently recognized that $b$ should depend on $z$ and $R$ (e.g.,
Fry 1996; Mo \& While 1996). As a matter of fact, it is more realistic to
abandon the simple linear biasing ansatz (\ref{eq:linbias}) completely,
and formulate the biasing in terms of the conditional probability
function $P(\deltag|\deltam)$ of $\deltag$ at a given $\deltam$. Then
equation (\ref{eq:linbias}) is rephrased as
\begin{equation}
\label{eq:linbias2}
\overline\deltag(\deltam) = \int_{-1}^{\infty} 
\deltag ~P(\deltag|\deltam) ~d\deltag .
\end{equation}
Obviously the relation between $\deltag$ and $\deltam$ is neither
linear nor deterministic.  This general concept -- nonlinear and
stochastic biasing -- was introduced and developed in a seminal paper
by Dekel \& Lahav (1999), which inspired numerous recent activities in
this field (e.g., Pen 1998; Taruya, Koyama \& Soda 1999; Blanton et
al. 1999; Matsubara 1999; Taruya 2000; Somerville et al. 2000; 
Taruya et al. 2000).

Then the crucial question is the physical interpretation of
$P(\deltag|\deltam)$.  There are (at least) two different
interpretations for its physical origin. The first is based on the
fact that $\deltag$ is meaningless unless one specifies many {\it
  hidden} variables characterizing the galaxies in the catalogue, for
instance, their luminosity, mass of dark matter and gas, temperature,
physical size, formation epoch and merging history, among others.
This list is already long enough to convince one for the inevitably
broad distribution of $P(\deltag|\deltam)$.  In this spirit, Blanton
et al.  (1999) proposed that the gas temperature of a local patch is
the important variable to control $P(\deltag|\deltam)$ on the basis of
cosmological hydrodynamical simulations.  The other adopts the view
that our universe is fully specified by the primordial density field
of dark matter. According to this interpretation, all the properties
of any galaxy should be in principle computable given an initial
distribution of dark matter field in the entire universe, at least in
a statistical sense.  Actually this is exactly what the cosmological
hydrodynamical simulations attempt to do. The gas temperature of a
local patch, for instance, should be determined by a non-local
attribute of the dark matter fluctuations. Clearly the above
interpretations are not conflicting, but rather stress the two
different aspects of the physics of galaxy formation which is poorly
understood at best.

In this paper, we present an analytical model for nonlinear stochastic
biasing by combining the above two interpretations in a sense.
Specifically we derive the joint probability function of $\deltah$ and
$\deltam$, $P(\deltah, \deltam)$, from the distributions of the
formation epoch and mass of halos on the basis of the extended
Press-Schechter theory. In contrast to previous work which were based
on the results of numerical simulations (Kravtsov \& Klypin 1999;
Blanton et al. 1999; Somerville et al. 2000), our model provides, for
the first time, an analytic expression for $P(\deltah, \deltam)$. Thus
one can compute the various biasing parameters in an arbitrary
cosmological model at a given redshift and a smoothing length in a
straightforward fashion.  We derive the joint probability function
assuming the primordial random-Gaussianity of the dark matter density
field, and thus the results are sufficiently general.  

We note here, however, that our primary purpose is to present a
general formulation to predict biasing properties of halos, and not to
make detailed predictions at this point. In fact we adopt a few
approximations with limited validity in presenting specific examples,
but this is not essential in our paper and the resulting predictions
can be improved in a straightforward manner if other
analytical/numerical approximations become available.  Nevertheless we
would like to emphasize that our simple analytical prescription
largely explains the basic features of the biasing parameters reported
in the previous numerical simulations (Kravtsov \& Klypin 1999;
Somerville et al. 2000). Thus our prescription is supposed to capture
the most important processes in the halo biasing.

We organize the paper as follows. In \S \ref{sec:formalism}, we describe
a general formalism for the one-point statistics of the galaxy and the
mass distributions from a point of view of the hidden variable
interpretation of the nonlinear and stochastic biasing theory.  Applying
this general formalism, we develop a model for dark matter halo biasing
treating the halo mass and formation epoch as the hidden variables in \S
\ref{sec:model}.  The resulting expression for the conditional
probability function, $P(\deltah|\deltam)$, can be numerically
evaluated using the extended Press-Schechter theory, and we show various
predictions for the halo biasing in \S \ref{sec:results}.  In particular,
we pay attention to their scale-dependence and redshift evolution, and
compare our model predictions to previous simulation results. Finally
section \ref{sec:conclusions} is devoted to the conclusions and
discussion.

\section{Formulation of Nonlinear Stochastic Biasing in Terms of
  Hidden Variables}
\label{sec:formalism}

In this section, we present a general formalism of biasing for the
one-point statistics of the galaxies and the mass smoothed with the
radius $R$.  While we specifically focus on the second order
statistics and discuss their nonlinearity and stochasticity in the
present paper, the formalism is readily applicable to higher-order
statistics.

\subsection{Probability distribution function and hidden variables}
\label{subsec:PDFinhidden}
Recall that the fluctuations of galaxies and the dark matter density
field are given by
\begin{equation}
  \label{density_field}
  \deltag({\bf x},z)=
  {n_{\rm gal}({\bf x},z)\over\bar{n}_{\rm gal}}-1, 
  \qquad
  \deltam({\bf x},z) =
  {\rho_{\rm mass}({\bf x},z) \over \bar\rho_{\rm mass}} - 1 ,
\end{equation}
where the variables with over-bar denote the homogeneous mean over the
entire universe. We evaluate these quantities smoothed with a
spherical symmetric filter function $W({\bf x};R)$:
\begin{equation}
  \label{eq:smoothed-delta}
  \delta({\bf x}, z|R)=\int d^{3}{\bf y}~
  W(|{\bf x}-{\bf y}|;R) \delta ({\bf y},z), 
\end{equation}
which corresponds to quantities in equation (\ref{eq:dgdm}).  The
one-point statistics of galaxies and the mass are calculated from
equation (\ref{eq:smoothed-delta}).

In general, fluctuations of the biased objects $\deltag$ are specified
by multi-variate functions of $\deltam$ and other observable and
unobservable variables, $\vec{\cal U}$, $\vec{\cal O}$, characterizing
the sample of objects. Then one can formally write 
\begin{equation}
    \label{delta-g}
    \deltag=\gal(R,z|\deltam,~\vec{\cal U},~\vec{\cal O}) ,
\end{equation}
where we use $\gal$ to denote the galaxy number density contrast at a
position ${\bf x}$ and redshift $z$ smoothed over a scale $R$ as a
function of $\deltam$, $\vec{\cal U}$, and $\vec{\cal O}$.  In the
above expression, $\vec{\cal O}$ and $\vec{\cal U}$ should be also
regarded as functions of (${\bf x},z)$ smoothed over the size $R$.  In
practice, galaxies in redshift surveys are identified and/or
classified according to their magnitude, spectral and morphological
type. The spatial clustering of galaxies should naturally depend on
those observable quantities, $\vec{\cal O}$. Since any sample of
galaxies is selected over a range of values for $\vec{\cal O}$, the
distribution of $\vec{\cal O}$ leads to the stochasticity of the
clustering biasing of the sample. Furthermore, the unobservable
quantities or the {\it hidden variables}, $\vec{\cal U}$, which
characterize an individual galaxy reflecting the different history of
gravitational clustering and merging, radiative cooling, and
environment effects, should provide additional stochasticity. Although
these processes could be related to the dark matter density
fluctuation in a ``non-local'' fashion, we intend to incorporate those
effects into our biasing model by a set of local functions such as the
gas temperature, mass of the hosting halos, and the formation epoch of
galaxies. 

While the distinction between $\vec{\cal O}$ and $\vec{\cal U}$ is
{\it conceptually} important, it may not be easy or straightforward in
reality. Nevertheless it is not essential in our prescription below as
long as their probability distribution function (PDF):
\begin{equation}
    \label{multi-jointPDF}
  P= P(\deltam,~\vec{\cal U}) 
\end{equation}
is specified. It should be noted that the above expression implicitly
depends on the smoothing radius and the redshift for the given classes
$\vec{\cal O}$. The statistical information of galaxy biasing is obtained
by averaging over the joint PDF $P(\deltam,\deltag)$. To be more
specific, the joint average of a function ${\cal F}(\deltam,\deltag)$
is defined by
\begin{equation}
  \label{ensemble-average2}
 \llangle{\cal F}(\deltam, \, \deltag) \rrangle = 
\int d\deltam \int d\deltag \, P(\deltam,\deltag) \,
  {\cal F}(\deltam,~\deltag),       
\end{equation}
where we use $\llangle\cdots\rrangle$ so as to explicitly denote the
joint average over the two stochastic variables, $\deltam$ and
$\deltag$.

In our prescription, however, it is more convenient to perform the
averaging over $P(\deltam,~\vec{\cal U})$ instead of
$P(\deltam,\deltag)$:
\begin{equation}
 \llangle{\cal F}(\deltam,~\deltag) \rrangle =
\int d\deltam \int\cdots\int \prod_{i} d\,{\cal U}_{i} \,
P(\deltam,~\vec{\cal U})~{\cal F}(\deltam,\,\gal) ,
  \label{ensemble-average}
\end{equation}
where the variable $\gal$ in the argument of ${\cal F}$ should be
regarded as a function of $\deltam$ and $\vec{\cal U}$
(eq.[\ref{delta-g}]). Of course the two expressions
(\ref{ensemble-average2}) and (\ref{ensemble-average}) should give the
identical result, and thus one obtains
\begin{equation}
  \label{PDF-hidden}
P(\deltam,\deltag)d\deltag=\int\cdots\int_{{\cal C}(\vec{\cal U})}
\prod_{i}d\,{\cal U}_{i}\,P(\deltam,\,\vec{\cal U}), 
\end{equation}
where the region ${\cal C}(\vec{\cal U})$ of integration is defined as
\begin{equation}
\label{constraint}
{\cal C}(\vec{\cal U})=\{\vec{\cal U}|~\deltag\leq \, \gal \, \leq
\deltag+d\deltag \}, 
\end{equation}
for a given $\deltam$. Equation (\ref{PDF-hidden}) implies that the
unobservable information represented by $\vec{\cal U}$ serves as a
source for stochasticity between $\deltam$ and $\deltag$ . In other
words, equations (\ref{PDF-hidden}) and (\ref{constraint}) can be
regarded as to the definition of the joint PDF $P(\deltam,\deltag)$.  
Once $P(\deltam,\vec{\cal U})$ is specified, equation (\ref{PDF-hidden})
can be computed numerically in a straightforward manner. In the next
section, we will develop a simple analytical model for the dark halo
biasing and explicitly calculate the joint PDF according to this
prescription assuming that the formation epoch of halos $\zf$ is the
major variable in $\vec{\cal U}$.

Finally when the joint PDF is given, it is straightforward to
calculate the conditional PDF of galaxies for a given overdensity
$\deltam$:
\begin{equation}
    \label{conditionalPDF}
    P(\deltag|\deltam)=\frac{P(\deltam,\deltag)}{P(\deltam)} , 
\end{equation}
where the one-point PDF of the mass $\deltam$ is related to 
\begin{equation}
    P(\deltam)=\int\cdots\int \prod_{i} d \, {\cal U}_i \, 
    P(\deltam, \,\vec{\cal U})
    =\int d\deltag \, P(\deltam,\deltag).
    \nonumber
\end{equation}

\subsection{Second-order statistics and the biasing parameters} 
\label{subsec:second-order}

Turn next to the statistical quantities characterizing the PDF. For
this purpose, we consider the second-order statistics, variance of
galaxies, variance of mass, and covariance of galaxies and mass, which
are defined \footnote{While we use $\llangle \cdots \rrangle$ (joint
  average over $\deltam$ and $\deltag$) even in the definition of
  $\sigma_{\rm gg}$ and $\sigma_{\rm mm}$, it can be replaced by the
  single average over $\deltag$ and $\deltam$, respectively.}
respectively as
\begin{equation}
\label{variance-covariance}
 \sigma^2_{\rm gg}(R,z)\equiv\llangle\deltag^2\rrangle ,
~~~~
 \sigma^2_{\rm mm}(R,z)\equiv\llangle\deltam^2\rrangle ,
~~~~
 \sigma^2_{\rm gm}(R,z)\equiv\llangle\deltag\deltam\rrangle .
\end{equation}
It should be also noted here that the last quantity, covariance of
galaxies and mass, is not positive definite unlike the first two.  As
we will see below, however, this is always positive for a range of
model parameters of cosmological interest which we survey. Thus we use
$\sigma_{\rm gm}(R,z) \equiv \llangle\deltag\deltam\rrangle^{1/2}$ for
the notational convenience. Their ratios represent the degree of the
biasing and stochasticity:
\begin{equation}
\label{eq:ratio1} 
 \bvar\equiv \frac{\sigmagg}{\sigmamm},
~~~~~~
 \rcorr\equiv\frac{\sigma^{2}_{\rm gm}}{\sqrt{\sigmagg\sigmamm}},    
\end{equation}
which are sometimes quoted as ``the'' biasing parameter and the cross
correlation coefficient (Dekel \& Lahav 1999). In this paper, we use
the subscripts, var and corr, for the above parameters in order to
avoid possible confusions with other parameters introduced in previous
papers.

The above definitions of $\bvar$ and $\rcorr$ do not yet fully
distinguish the nonlinear and stochastic nature of the biasing in a
clear manner. Thus we introduce more convenient statistical measures,
$\nonl$ and $\scatt$, which quantify the two effects separately.  For
this purpose, the conditional mean of $\deltag$ for a given $\deltam$
(Dekel \& Lahav 1999):
\begin{equation}
    \label{eq:conditional_mean}
    \overline\deltag(\deltam)=
    \int d\deltag \, P(\deltag|\deltam) \, \deltag
\end{equation}
plays a key role.  Note that the average of
$\overline\deltag(\deltam)$ over $\deltam$ vanishes from definition
(\ref{density_field}) :
\begin{eqnarray}
  \int d\deltam\, P(\deltam) \, \overline\deltag(\deltam) &=&
\int d\deltag \left[\int d\deltam\,P(\deltam,\deltag)\right] \deltag
\cr
&=& \int d\deltag  \, P(\deltag) \, \deltag
= 0 .
\end{eqnarray}

The nonlinearity of biasing refers to the departure from the linear
proportional relation between $\overline\deltag$ and $\deltam$.
This can be best quantified by the following measure:
\begin{equation}
    \label{eq:nonl}
\nonl^2 \equiv \frac{\llangle\deltam^2\rrangle 
              \, \llangle\overline\deltag^2\rrangle}
    {\llangle\deltam\overline\deltag\rrangle^2} - 1
= \frac{\sigma^2_{\rm mm} \, \llangle\overline\deltag^2\rrangle}
{\sigma^4_{\rm gm}} - 1 .
\end{equation}
The second equality in the above comes from the fact that
$\llangle\deltam\overline\deltag\rrangle =
\llangle\deltam\deltag\rrangle$.  From the Schwartz inequality,
one show that the right-hand-side of the above equation is
non-negative and vanishes only if the linear coefficient $b_{1} \equiv
\overline\deltag/\deltam$ is independent of $\deltam$.

The stochasticity of biasing corresponds to the scatter or dispersion
of $\deltag$ around its conditional mean $\overline\deltag(\deltam)$.
Averaging this scatter over $\deltam$ with proper normalization, we
define the following measure for the stochasticity of biasing:
\begin{equation}
    \label{eq:scatt}
\scatt^2 \equiv \frac{\llangle\deltam^{2}\rrangle \,
      \llangle(\deltag-\overline\deltag)^2\rrangle}
    {\llangle\deltam\overline\deltag\rrangle^2} 
= \frac{\sigma^2_{\rm mm} \, 
[\sigma^2_{\rm gg} - \llangle\overline\deltag^2\rrangle]}
{\sigma^4_{\rm gm}} .
\end{equation}
Since the galaxy density field $\deltag$ (eq.[\ref{delta-g}]) depends
on many variables other than $\deltam$, $\scatt$ does not vanish in
general. In turn, $\scatt=0$ corresponds to the unlikely case that
$\deltag$ is uniquely determined by $\deltam$ and thus $\deltag
=\overline\deltag(\deltam)$.

The galaxy biasing still exists even when $\scatt=\nonl=0$. In fact, a
simple linear and deterministic biasing (\ref{eq:linbias}) falls into
this category. This effect can be separated out from the covariance or
linear regression of $\deltag$ and $\deltam$ (Dekel \& Lahav 1999) as
follows:
\begin{equation}
    \label{eq:bcov}
\bcov \equiv  \frac{\llangle\deltam\overline\deltag\rrangle}
    {\llangle\deltam^2\rrangle}
= \frac{\sigma^2_{\rm gm}}{\sigma^2_{\rm mm}} .
\end{equation}
This quantity is equivalent to the coefficient of the leading order in
the Taylor expansion, $\overline\deltag=\bcov\deltam+\cdots$, in a
perturbative regime, $\deltam \ll 1$ (Taruya \& Soda 1999).

The biasing parameters that we introduced are related to the more
conventional biasing coefficients (eq.[\ref{eq:ratio1}]) as
\begin{equation}
\label{eq:ratio2}
 \bvar= \bcov (1+\scatt^2+\nonl^2)^{1/2},
~~~~~~~
\rcorr= \frac{1}{\sqrt{1+\scatt^2+\nonl^2}} . 
\end{equation}
These relations clearly indicate that $\scatt$ and $\nonl$ separate
the stochastic and nonlinear effects which are somewhat degenerate in
the definitions of $\bvar$ and $\rcorr$.  Also it may be useful to
express our biasing parameters in terms of those introduced by Dekel
\& Lahav (1999) :
\begin{equation}
 \bcov=\bhat,~~~~~\scatt=\frac{\sigmab}{\bhat},~~~~~
\nonl=\sqrt{\left(\frac{\btilde}{\bhat}\right)^2-1} . 
\end{equation}
Of course the two sets of choice are essentially equivalent, we hope
that our notation characterizes the physical meaning of nonlinear and
stochastic biasing more clearly.  Finally, while the present paper is
focused on the analyses of the second-order statistics, 
it is fairly straightforward to extend the above formalism to the
higher-order statistics.

\section{An Analytic Model of Nonlinear Stochastic Biasing 
for Dark Halos \\ from the Formation Epoch Distribution}
\label{sec:model}

\subsection{Schematic picture of our halo biasing model 
\label{subsec:picture}}

The previous section describes the general formalism for the nonlinear
stochastic biasing in terms of the hidden variable interpretation. In
this section we present a specific model for halo biasing and discuss its
predictions according to the general formalism.  Before proceeding to
the technical details, it is useful to explain first the basic
picture of our model in a qualitative manner.

\begin{figure}
\begin{center}
   \leavevmode \epsfxsize=16cm \epsfbox{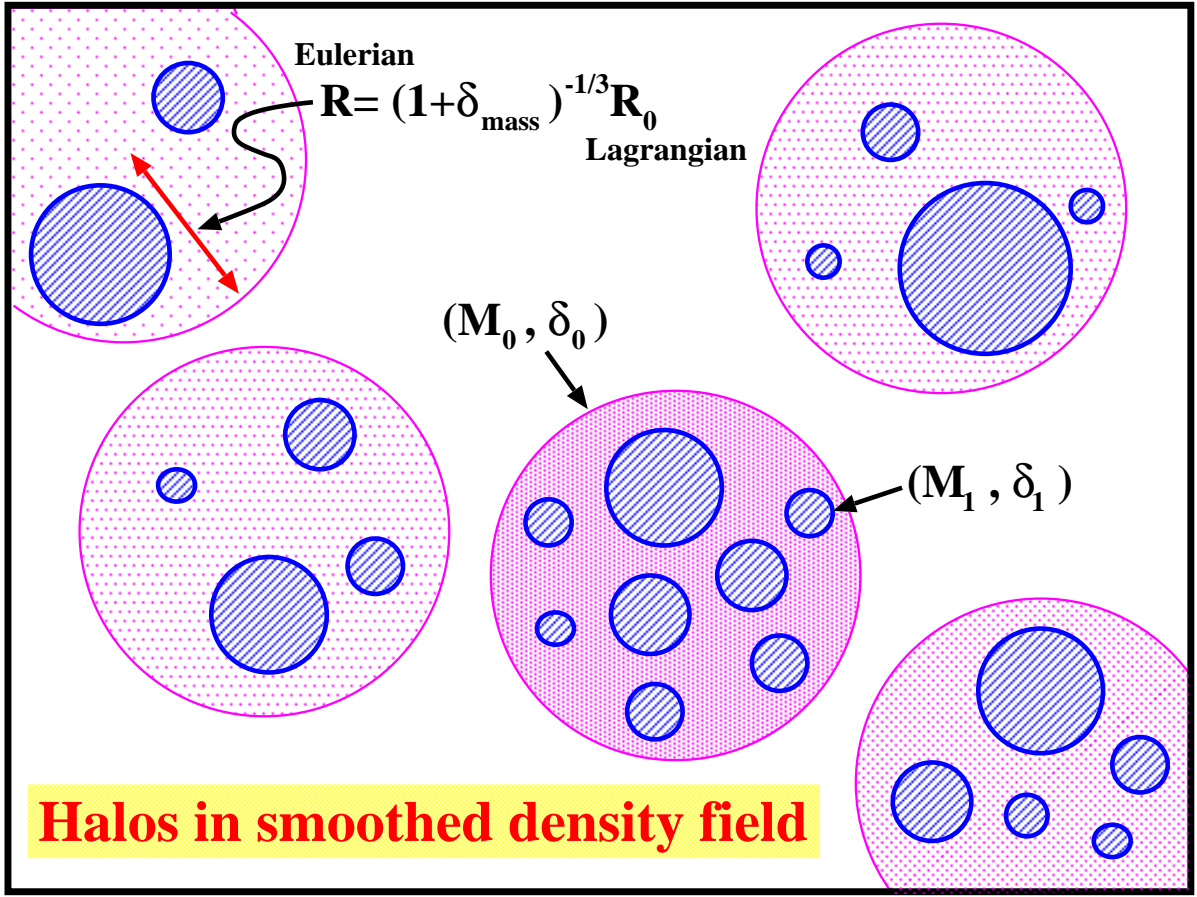}
\end{center}
\figcaption{ A schematic picture of our halo biasing model at a given
  redshift $z$.  The mass and halo density fields are top-hat smoothed
  with the {\it Eulerian} proper radius $R$, and $R_0$ is its
  Lagrangian counterpart.  Each sampling sphere is characterized by
  $R$ and the Eulerian mass density contrast $\deltam$, or
  equivalently by the enclosed mass $M_0$ and the linearly
  extrapolated mass density contrast $\delta_0$.  Within each sampling
  sphere, there are a number of dark halos characterized by their mass
  $M_1$ and the linearly extrapolated mass density contrast
  $\delta_1$. With a given $z$, $\delta_1$ is related to the formation
  epoch $\zf$ of the halo.
\label{fig:PS_halos} }
\end{figure}

As illustrated schematically in Figure \ref{fig:PS_halos}, we consider
the mass and galaxy density fields at redshift $z$ smoothed over the
top-hat {\it Eulerian} proper radius $R$. The mass density contrast
$\deltam$ computed in the Eulerian coordinate relates $R$ with its
Lagrangian coordinate counterpart $R_0 = (1+\deltam)^{1/3}R$ assuming
the spherical collapse. Then the mass in the sphere $M_0$ is simply
given by $(4\pi/3)\bar{\rho}_{\rm mass}R_0^3$, where $\bar{\rho}_{\rm
  mass}(z)$ is the physical mass density at $z$. Also the linearly
extrapolated mass density contrast $\delta_0$ in the sphere can be
evaluated from $\deltam$ on the basis of the nonlinear spherical
collapse model (e.g., Mo \& White 1996).

Each sphere of the Eulerian radius $R$ should contain a number of
gravitationally virialized objects i.e., {\it dark matter halos}.
Given $M_0$ and $\delta_0$, their conditional mass function can be
predicted by extended Press-Schechter theory (e.g., Bower 1991; Bond
et al. 1991).  Such halos are conventionally characterized by their
mass $M_1$ and linearly extrapolated mass density contrast $\delta_1$
assuming that their formation epoch is equivalent to the current
redshift $z$.  Kitayama \& Suto (1996a) pointed out that this
approximation often significantly changes the predictions for X-ray
cluster abundances on the basis of the Press-Schechter theory, and
proposed a phenomenological prescription to compute the formation
epoch $\zf$. In fact, the halo biasing derived by Mo \& White (1996)
is fairly sensitive to the difference of $z$ and $\zf$ as noticed by
Kravtsov \& Klypin (1999). Thus we extend the biasing model of Mo and
White (1996), in which $\zf$ should be specified {\it a priori}, by
considering explicitly the dependence on $\zf$ and averaging according
to the formation epoch distribution function of Lacey \& Cole (1993)
and Kitayama \& Suto (1996b).  This is a major and important
improvement of our model over the original proposal of Mo and White
(1996).  In our model, therefore, the formation epoch $\zf$ and the
mass of halo $M_1$ constitute the {\it hidden} variables $\vec{\cal
  U}$ (\S 2), and their PDFs generate the nonlinear and stochastic
behavior in the resultant halo biasing.

\subsection{Halo biasing from the extended Press-Schechter theory}
\label{subsec:EPStheory}

In what follows, we assume that the primordial mass density field
obeys the random - Gaussian statistics (e.g., Bardeen et al. 1986).
In this case, the (unconditional) mass function of dark halos (Press
\& Schechter 1974) :
\begin{equation}
\label{universalMF}
 n(M_{1},z;\delta_{1})dM_{1}=\frac{1}{\sqrt{2\pi}}
 \frac{\bar{\rho}_{\rm mass}}{M_{1}}
  \frac{\delta_{1}}{\sigma^3(M_{1},z)}
  \exp{\left[-\frac{\delta_{1}^2}{2\sigma^2(M_{1},z)}\right]}
\left|\frac{d\sigma^2(M_{1},z)}{dM_{1}}\right|dM_{1} 
\end{equation}
proves to be in reasonable agreement with results from $N$-body
simulations (e.g., Efstathiou et al. 1988; Suto 1993; Lacey \& Cole
1993,1994). Equation (\ref{universalMF}) corresponds to the comoving
number density of halos exceeding the critical density threshold
$\delta_1$ and of mass between $M_{1}$ and $M_{1}+dM_{1}$.  The value
for $\delta_{1}$ will be specified when we consider the one-point
function or the conditional PDF of the dark halos below (see eqs.
[\ref{eq:delta-halo}] and [\ref{weight}]).  The mass variance
$\sigma^2(M_1,z)$ is defined from the {\it linear} power spectrum of
mass density fluctuations $P_{\rm lin}(k)$ at present ($z=0$):
\begin{equation}
  \label{mass-variance}
 \sigma^2(M_1,z)=D^2(z)~\int\frac{dk}{2\pi^2}~k^{2}P_{\rm lin}(k)W^2(kR_1) ,
\end{equation}
where $D(z)$ is the linear growth factor normalized to unity at $z=0$ and   
the top-hat window function: 
\begin{eqnarray}
\label{top-hat}
 W(x)= \frac{3}{x^3}(\sin{x}-x\cos{x})
\end{eqnarray}
with the Lagrangian radius $R_1 \equiv (4\pi \bar{\rho}_{\rm
  mass}/3M_1)^{-1/3}$ is adopted.

Since our one-point statistics of halos is evaluated within a sample
of spheres of the Eulerian radius $R$, we need the conditional mass
function for halos within a sphere. For this purpose, we use the
extended Press-Schechter theory which predicts the conditional mass
function for halos of $(M_1,~\delta_1)$ and $(M_1+dM_1,~\delta_1)$ in
the background region of $(M_0,~\delta_0)$ as
\begin{eqnarray}
n(M_1,\delta_1|M_0,\delta_0;z)dM_1 &=&
  \frac{1}{\sqrt{2\pi}}
\frac{\bar{\rho}_{\rm mass}}{M_1}
\frac{\delta_1-\delta_0}{[\sigma^2(M_1,z)-\sigma^2(M_0,z)]^{3/2}}
\cr
&\times&  \exp{\left[-\frac{(\delta_1-\delta_0)^2}
        {2\{\sigma^2(M_1,z)-\sigma^2(M_0,z)\}}\right]}
\left|\frac{d\sigma^2(M_1,z)}{dM_1}\right|dM_1 .
\label{conditionalMF}
\end{eqnarray}

Then the biased density field for halos of mass $M_1$, which formed at
$\zf$ and are observed at $z$ within the sampling sphere of
$(R,\deltam)$ in Eulerian coordinates, is derived by Mo \& White (1996):
\begin{equation}
\label{eq:delta-halo}
\deltah = \halo(R,z|~\deltam,~\zf,~M_1) \equiv
(1+\deltam)\frac{n(M_1,~\delta_c(z,\zf)~|M_0,~\delta_0~;z)} 
{n(M_1,z;\delta_c(z,\zf))} - 1 .
\end{equation}
In the above, the critical threshold $\delta_c$ for those halos is
given as
\begin{equation}
\label{delc}
 \delta_c(z,\zf)=\delta_{c,0}~\frac{D(z)}{D(\zf)}, \qquad
 \delta_{c,0}\equiv \frac{3(12\pi)^{2/3}}{20}\simeq 1.69 ,
\end{equation}
again on the basis of the spherical collapse model. The remaining task
is to compute the mass, $M_0$, and the {\it linearly extrapolated}
mass density contrast, $\delta_0$, of the sampling sphere from its
Eulerian radius and density contrast $(R,\deltam)$. Since $\delta_0
\ll 1$, the former is simply given as
\begin{equation}
M_0 = {4\pi \over 3} \, \bar{\rho}_{\rm mass}(z)\,(1+\delta_0) \, R_0^3
\simeq {4\pi \over 3} \, \bar{\rho}_{\rm mass}(z) \, (1+\deltam)\, R^3 ,
\end{equation}
with $\bar{\rho}_{\rm mass}(z)$ being the physical mass density at
$z$. Finally we adopt the following fitting formula obtained in the
spherical collapse model (Mo \& While 1996) to compute $\delta_0$ in
terms of $\deltam$:
\begin{equation}
  \label{del0-del}
 \delta_0=\delta_{c,0}-\frac{1.35}{(1+\deltam)^{2/3}}
  +\frac{0.78785}{(1+\deltam)^{0.58661}}
   -\frac{1.12431}{(1+\deltam)^{1/2}} .
\end{equation}
While equations (\ref{delc}) and (\ref{del0-del}) were originally
derived in the Einstein-de Sitter model, we numerically checked that
they provide sufficiently accurate approximations for our present
purpose even in the $\Omega_0=0.3$ and $\lambda_0=0.7$ model that we
consider later. Thus we use the expressions (\ref{delc}) and
(\ref{del0-del}) irrespectively of the cosmological models in the
subsequent analysis.

Figure \ref{fig:dhdm} illustrates the dependence of $\halo$ on $M$ and
$\zf$ as a function of $\deltam$ smoothed over $R=8\himpc$ at $z=0$.
Given a halo mass, $\halo$ is very sensitive to the formation epoch
$\zf$ especially in the range of $\zf \simlt z+1$.  As $\zf$
increases, the dependence $\halo$ and $M$ and $\deltam$ becomes
significant; this reflects the fact that the larger mass halos
preferentially form in the denser regions than the average since the
typical halo mass that can be collapsed and virialized decreases at
higher $\zf$. Such $M$ and $\zf$ dependence of $\halo$ convolved with
the PDF of $M$ and $\zf$ leads to the nonlinear stochastic behavior of
the biasing of dark halos.

While we regard halo mass $M_1$ and its formation epoch $\zf$ as the
{\it hidden variables} in equation (\ref{eq:delta-halo}), they may not be
entirely unobservable. One may infer the halo mass and the formation
epoch for an individual galaxy by combing the observed luminosity,
color and metallicity with a galaxy evolution model. In this case,
their probability distribution functions need to be convolved with
such observational selection functions with our prior distribution.
Except for this correction, our methodology presented below remains
the same.

\subsection{Formation epoch distribution of dark halos}
\label{subsec:formation epoch}

As indicated in equation (\ref{eq:delta-halo}), the amplitude of halo
biasing is explicitly dependent on its formation epoch $\zf$.  Thus
the simple approximation of $\zf=z$ may lead to even qualitatively
incorrect predictions for the biasing. In fact this was shown to be
the case in recent N-body simulations by Kravtsov \& Klypin (1999).
The importance of the distribution of $\zf$ has been emphasized by
Kitayama \& Suto (1996a) in a different context, and a model for its
PDF was proposed by Lacey \& Cole (1993). Incidentally Catelan et al.
(1998a,b) also proposed a different model of halo biasing considering
the $\zf$-dependence. Their model simply treats $\zf$ as a free
parameter and does not properly take account of its distribution
function. Our model presented here incorporates the distribution
function of the formation epoch explicitly.

Adopting the excursion set approach (Bond et al. 1991) and {\it
  defining} the formation redshift $\zf$ of a particular halo of mass
$M$ at $z$ as the epoch when the mass of its most massive progenitor
exceeds $M/2$ for the first time, Lacey \& Cole (1993) derived the
differential distribution of the halo formation epoch $\partial
p/\partial \zf$. Their result is expressed as
\begin{equation}
 \frac{\partial p}{\partial \zf}(\zf~|M,~z)d\zf=
\frac{\partial p}{\partial \tilde{\omega}_{\rm f}}
(\tilde{\omega}_{\rm f},~M)~
\frac{\partial\tilde{\omega}_{\rm f}}{\partial \zf}d\zf,
\label{dp-dzf}
\end{equation}
\begin{equation}
 \frac{\partial p}{\partial \tilde{\omega}_{\rm f}}
 (\tilde{\omega}_{\rm f},~M)=
 -\frac{1}{\sqrt{2\pi}}\int_0^1 \frac{d\tilde{S}}{\tilde{S}^{3/2}}
 \frac{M}{M'(\tilde{S})}\left(1-\frac{\tilde{\omega}_{\rm f}^2}
  {\tilde{S}}\right)e^{-\tilde{\omega}_{\rm f}^2/(2\tilde{S})},
\label{dp-domegaf}
\end{equation}
where 
\begin{eqnarray}
 \tilde{\omega}_{\rm f}(\zf,M,z) &\equiv&
 \frac{\delta_{c}(z,\zf)-\delta_{c,0}}
  {\sqrt{\sigma^2(M/2,z)-\sigma^2(M,z)}},
\cr
 \tilde{S}(M',M) &\equiv& 
 \frac{\sigma^2(M',z)-\sigma^2(M,z)}
  {\sigma^2(M/2,z)-\sigma^2(M,z)} .
\label{ss}
\end{eqnarray}
The function $M'(\tilde{S})$ in the integrand of equation
(\ref{dp-domegaf}) can be obtained from equation (\ref{ss}).  While
the above expressions are rather complicated, practical fitting
formulae for the mass variance in cold dark matter (CDM) models and
for the formation epoch distribution were obtained in Kitayama \& Suto
(1996b) which we adopt throughout the analysis. Those are summarized
in Appendices A and B for convenience.

It should be noted that the definition of halo formation is somewhat
ambiguous in the framework of the extended PS theory. This aspect is
explored in Kitayama \& Suto (1996a), and their Figure 1
explicitly shows how the result is dependent on the adopted ratio of
the current halo mass and the progenitor mass at the formation epoch.
The figure implies that the resulting formation rate is fairly
insensitive to the value around 0.5 that we adopt here. 

Figure \ref{fig:dpdzf} plots the formation epoch distribution for
halos selected at $z$ in CDM models. Specifically we choose
$(\Omega_{0},\lambda_{0},\sigma_{8},h)$ $=(0.7,0.3,1.0,0.7)$ (Lambda
CDM; hereafter LCDM ) and $(1.0,0.0,0.6,0.5)$ (Standard CDM; hereafter
SCDM). The top-hat mass fluctuation amplitude at $8\himpc$,
$\sigma_8$, is normalized to the cluster abundance (Kitayama \& Suto
1997).  We show results for halos of $M=10^{11}h^{-1}\msun$(dashed), 
$10^{12}\msun$(solid) and $10^{13}h^{-1}\msun$(dot-dashed) 
in LCDM, and $10^{12}\msun$(dotted) in SCDM.  The
shape of those distribution functions is quite similar, and
characterized by a peak around $\zf = z+ (0.5\sim1)$. The peak
redshift becomes closer to the observed one, $z$, and the distribution
around the peak becomes narrower as the halo mass increases, both of
which are easily understood in the hierarchical clustering picture
like the present models.

Note that the SCDM model generally predicts a more sharply peaked
distribution closer to $z$ than the LCDM model (compare solid and
dotted lines in Fig.\ref{fig:dpdzf}). This is also reasonable from the
fact that the growth of fluctuations is rapid in SCDM and thus halos
form only recently. Thus the formation epoch distribution is fairly
sensitive to the cosmological parameters.

\subsection{Conditional and joint probability distribution functions 
  of dark halos and mass \label{subsec:pdfs} }

Now we are in a position to explicitly construct the conditional
probability distribution of the dark halo $P(\deltah|\deltam)$ for a
given $\deltam$, and the joint probability distribution $P(\deltah,
\deltam)$. Basically we follow the prescription described in \S
\ref{subsec:PDFinhidden}, but in slightly different order.

We first compute $P(\deltah|\deltam)$ applying equation
(\ref{PDF-hidden}):
\begin{equation}
    P(\deltah|\deltam) ~d\deltah~=~{\cal N}^{-1}~
    \int\int_{{\cal C}(M,\zf)}dM~ d\zf~~
\frac{\partial p}{\partial \zf}(z_{f}|M,z)~n(M,z;\delta_{c,0}) ,
\label{weight}
\end{equation}
where the region ${\cal C}(M,\zf)$ of the integration is determined
from the following conditions:
\begin{eqnarray}
    \label{eq:constraint}
{\cal C}(M,\zf) = \{~ (M,\zf)~|~ 
     \deltah\leq~\halo(R,z|\deltam,M,\zf)~\leq\deltah+d\deltah , \cr
 M_{\rm min}\leq M \leq M_{\rm max}, \quad z\leq \zf \leq \infty ~\}
\hspace*{1cm}
\end{eqnarray}
The normalization factor ${\cal N}$ is defined as
\begin{equation}
    \label{eq:normalization}
{\cal N}=
    \int_{M_{\rm min}}^{M_{\rm max}}dM \int_{z}^{\infty} d\zf~~
    \frac{\partial p}{\partial \zf}(\zf|M,z)~n(M,z;\delta_{c,0}) . 
\end{equation}
The integrand in equation (\ref{weight}) just corresponds to the joint
PDF $P(\vec{\cal U}|\deltam)$ for a given $\deltam$. 

Since the joint PDF is simply computed according to
\begin{equation}
\label{eq:jointpdf}
P(\deltam,\deltah) = P(\deltah|\deltam) \, P(\deltam) ,
\end{equation}
one needs a reliable model for the one-point PDF of dark matter
density contrast, $P(\deltam)$. Fortunately it has been known that
this can be empirically approximated by the log-normal distribution
function to a good accuracy (e.g., Coles \& Jones 1991; Kofman et al.
1994; Bernardeau \& Kofman 1995; Taylor \& Watts 2000):
\begin{equation}
    \label{lognormalPDF}
    P(\deltam)\,d\deltam=\frac{1}{\sqrt{2\pi}\sigma_1}
    \exp\left[-\frac{\left\{\ln(1+\deltam)
          +\sigma_{1}^{2}/2\right\}^{2}}
      {2\sigma_{1}^{2}}\right]\frac{d\deltam}{1+\deltam} ,
\end{equation}
where
\begin{equation}
    \sigma_1^{2}=\ln(1+\sigma_{\rm mm}^{2}) ,
\end{equation}
and $\sigmamm$ is defined in equation (\ref{variance-covariance}).
Note that equation (\ref{lognormalPDF}) reduces to the Gaussian
distribution for $\sigmamm \ll 1$, and thus this model again assumes
the primordial random-Gaussian density field implicitly as our entire
analysis. Finally we adopt the fitting formula (Peacock \& Dodds 1996)
for the nonlinear CDM power spectrum $P_{\rm nl}(k,z)$ in computing
the mass variance:
\begin{equation}
      \label{eq:peacock-dodds}
      \sigma_{\rm mm}^{2}(R,z)=\int
      \frac{dk}{2\pi^{2}}~k^{2}P_{\rm nl}(k,z)W^{2}(kR)~,  
\end{equation}
with $W(kR)$ being the top-hat smoothing function
(eq.[\ref{top-hat}]).  The validity of the log-normal approximation
for the one-point PDF is examined by Bernardeau \& Kofman (1995);
their Figure 10 indicates that equation (\ref{lognormalPDF})
reproduces the simulation results very accurately at least for $R
\simgt 5h^{-1}$Mpc. Although the accuracy on smaller scales is not
shown quantitatively, it would be reasonable to assume that the
approximation is acceptable up to $R\sim 1h^{-1}$Mpc. Also our
statistical results are not sensitive to the tail of such PDF in any
case.

Substituting the analytical expressions for ${\partial p}/{\partial
  \zf}(z_{f}|M,z)$ and $n(M,z;\delta_{c,0})$ discussed in \S
\ref{subsec:formation epoch} into equation (\ref{weight}), one may
numerically compute the conditional PDF $P(\deltah|\deltam)$, and 
the joint PDF $P(\deltah,\deltam)$ from equation (\ref{eq:jointpdf}).
Then all the statistical quantities can be evaluated using equation
(\ref{ensemble-average2}).  In practice, however, it is more
convenient and even accurate to use (\ref{ensemble-average}) which in
the present case is written explicitly as
\begin{eqnarray}
\label{average}
    \llangle{\cal F}(\deltah,~\deltam)\rrangle &=& 
  {\cal N}^{-1}~\int_{-1}^{\infty}d\deltam ~ P(\deltam) \cr
&\times&
\left[\int_{M_{\rm min}}^{M_{\rm max}}dM \int_{z}^{\infty} dz_{f}~
{\cal F}(\halo,~\deltam)~\frac{\partial p}{\partial z_{f}}(z_{f}|M,z)~
    n(M,z;\delta_{c,0})~ \right] .
\end{eqnarray}
We use the above expression in evaluating the various biasing parameters
below except in presenting the PDFs directly.

\section{Results in CDM models}
\label{sec:results}

We present several specific predictions applying our nonlinear
stochastic halo biasing model to representative CDM models mentioned in
\S \ref{subsec:formation epoch}.  Throughout the analyses, we consider
the range of halo mass between $M_{\rm min}=10^{11}h^{-1}\msun$ and
$M_{\rm max}=10^{13}h^{-1}\msun$ unless otherwise stated.

The general formulation described in the previous section should work
in principle even on fully nonlinear scales.  In practice, however,
the results presented below are limited by the halo exclusion effect
(due to the finite size of halos) and our approximation, equation
(\ref{lognormalPDF}), for the one-point PDF. The validity of both
effects should be carefully checked on small scales.  Since a typical
virial radius of a halo of mass $M$ in LCDM model is $\sim
0.5(M/10^{13}M_\odot)^{1/3} h^{-1}$ Mpc, the exclusion effect cannot
be neglected below $R=1h^{-1}$Mpc but is not so strong for $R \simgt
3h^{-1}$Mpc even for our largest mass considered ($M_{\rm
  max}=10^{13}h^{-1}\msun$). Also the validity of of the log-normal
approximation is already remarked in subsection 3.4.  Thus we expect
that our predictions below are fairly reliable up to $R \sim
3h^{-1}$Mpc.

\subsection{Conditional probability distribution of $\deltam$ and $\deltah$}
\label{subsec:condPDF}

Since the conditional PDF $P(\deltah|\deltam)$ plays a central role in
the Dekel \& Lahav (1999) description of the nonlinear stochastic
biasing, we first present $P(\deltah|\deltam)$ predicted from our
model. For this purpose, we start with equations (\ref{weight}) and
(\ref{eq:constraint}). Specifically we divide the $(\deltag, \deltam)$
plane in a $3000\times3000$ mesh, and accumulate the integrand of
equation (\ref{weight}) satisfying the constraint
(\ref{eq:constraint}) on each grid. The resulting PDFs are plotted in
Figure \ref{fig:prob_dh} for a given mass density contrast;
$\deltam=0$ (solid), $\deltam=1$ (dotted), $\deltam=2$
(dashed) and $\deltam=3$ (dot-dashed).  The upper and
lowers panels show the results at $z=0$ and $z=1$, respectively, with
the top-hat smoothing radius of $R=8h^{-1}$Mpc ({\it left}) and
$R=20h^{-1}$Mpc ({\it right}).

\begin{figure}
\begin{center}
   \leavevmode \epsfxsize=16cm \epsfbox{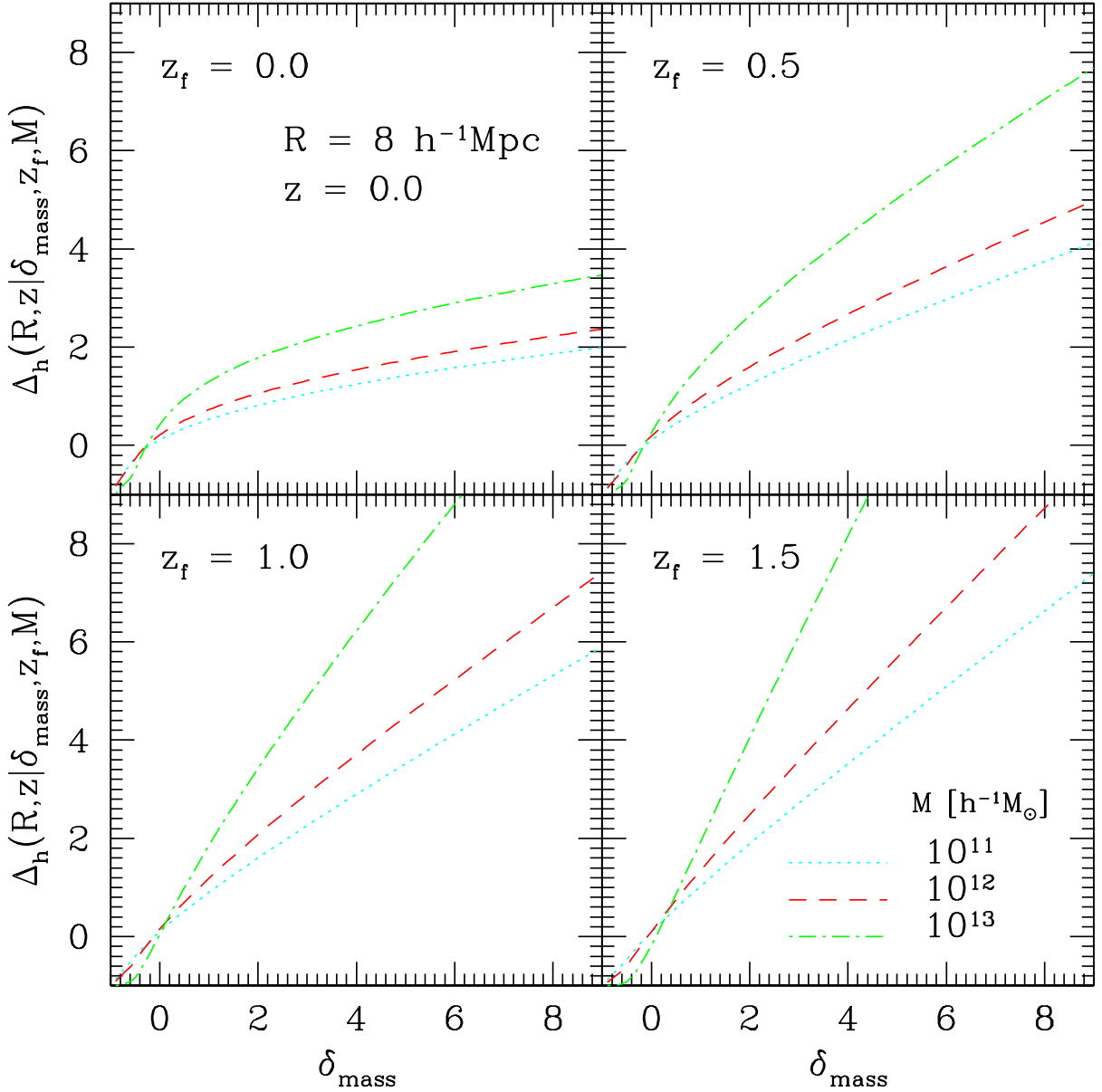}
\end{center}
\figcaption{The halo biasing field smoothed over $R=8\himpc$ at $z=0$ as
  a function of its mass $M$ and formation redshift $\zf$ in dotted
  ($M=10^{11}h^{-1}M_\odot$), dashed ($M=10^{12}h^{-1}M_\odot$),
  and dot-dashed ($M=10^{13}h^{-1}M_\odot$) lines.  {\it Upper-left:}
  $\zf=0$; {\it Upper-right:} $\zf=0.5$; {\it Lower-left:} $\zf=1.0$;
  {\it Lower-right:} $\zf=1.5$. 
\label{fig:dhdm} }
\end{figure}

\begin{figure}
\begin{center}
   \leavevmode \epsfxsize=15cm \epsfbox{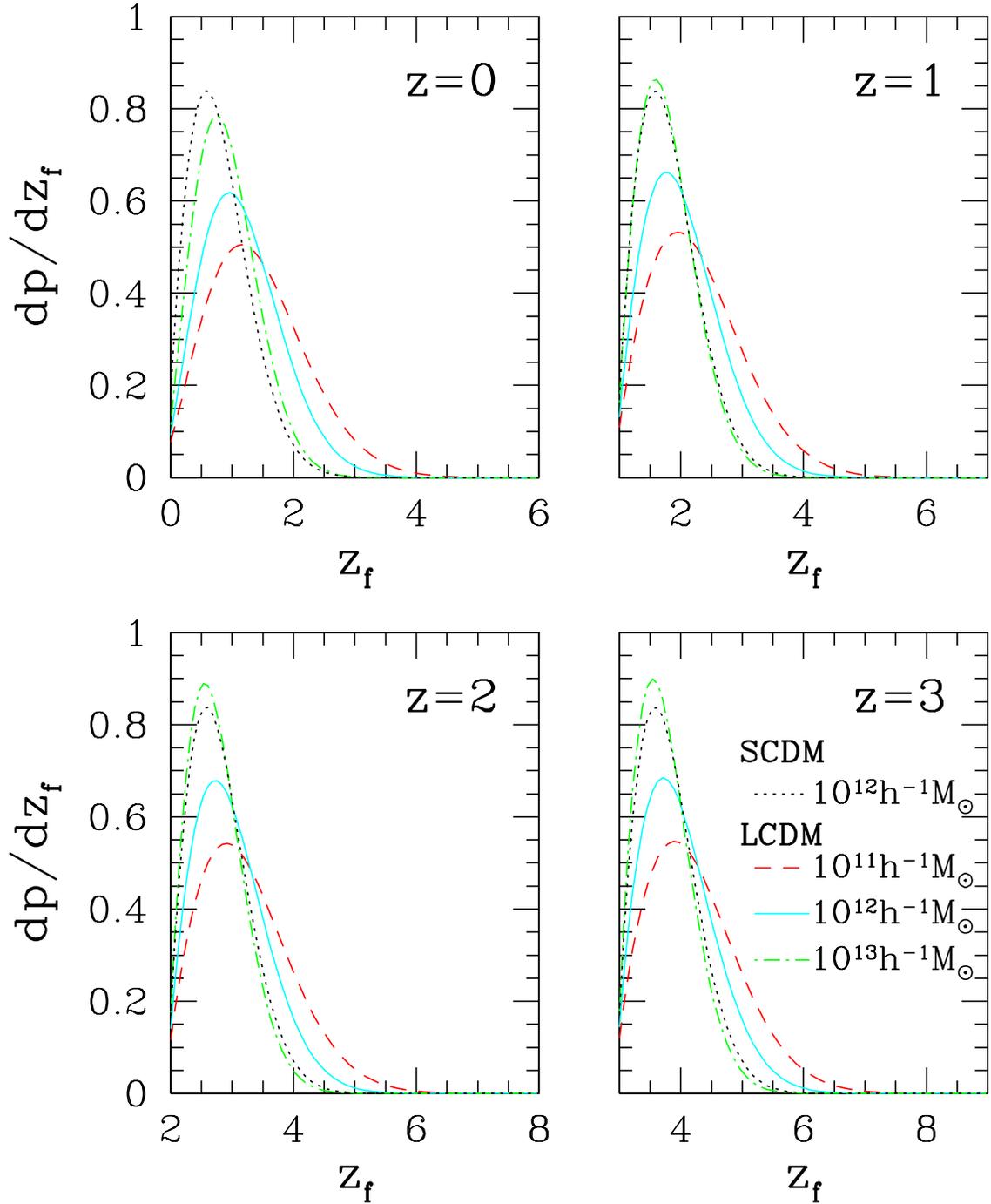}
\end{center}
\figcaption{Formation epoch distribution of dark halos.  
  Dotted lines indicate the results for $M=10^{12}h^{-1}M_\odot$
  in SCDM model, while dashed, solid and dot-dashed lines are respectively 
  for $M=10^{11}h^{-1}M_\odot$, $10^{12}h^{-1}M_\odot$ and 
  $10^{13}h^{-1}M_\odot$ in LCDM model. 
  {\it Upper-left:} $z=0$; {\it Upper-right:}
  $z=1$; {\it Lower-left:} $z=2$; {\it Lower-right:} $z=3$.
\label{fig:dpdzf} }
\end{figure}

\begin{figure}
\begin{center}
   \leavevmode\epsfxsize=15cm \epsfbox{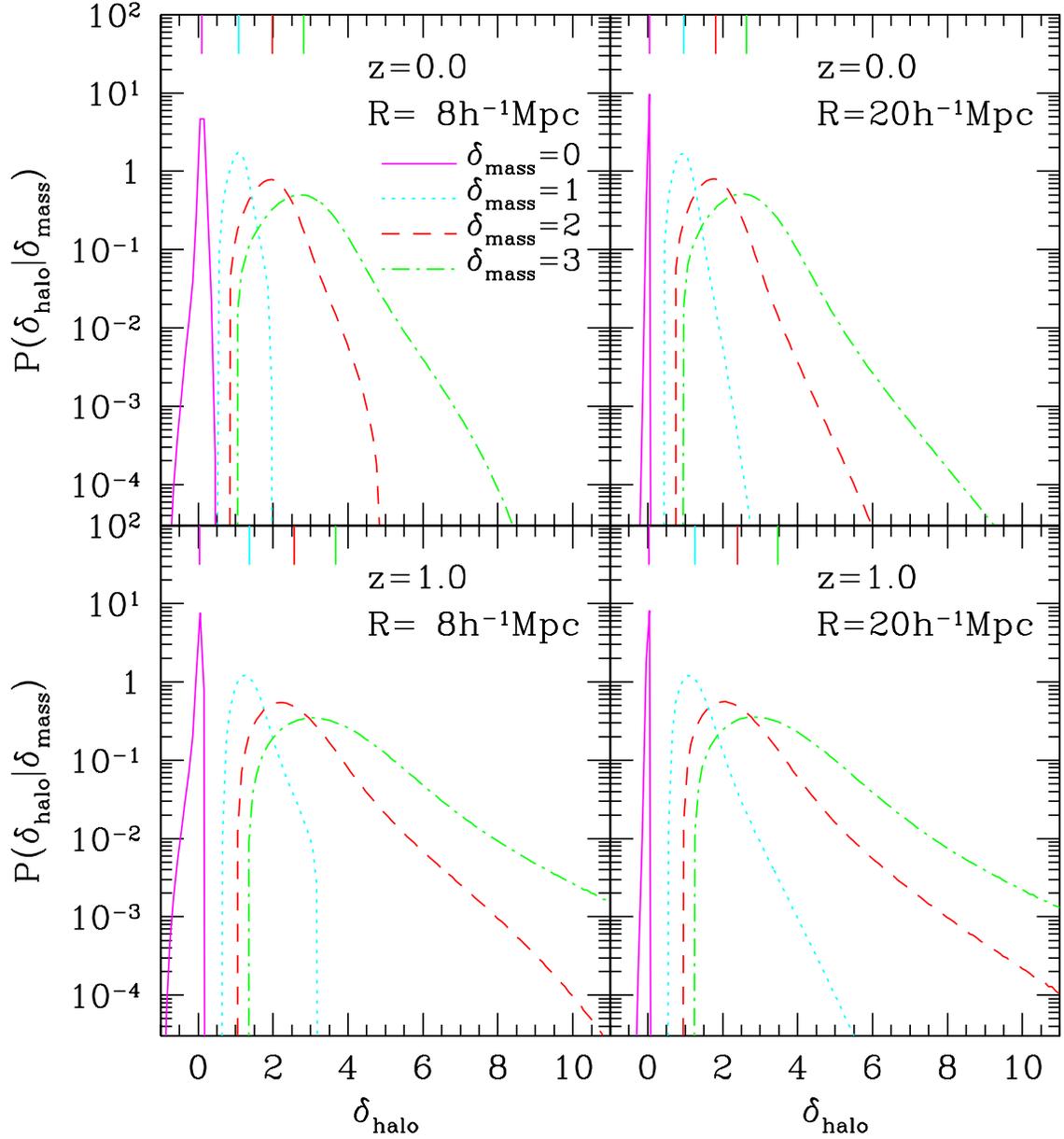}
\end{center}
\figcaption{Conditional probability of $\deltah$ for $\deltam=0$ 
  (solid), $1$ (dotted), $2$ (dashed) and $3$ (dot-dashed) 
  in LCDM model for $10^{11}h^{-1}M_\odot < M <
  10^{13}h^{-1}M_\odot$.  {\it Upper-left:} $z=0$ and $R=8h^{-1}$Mpc;
  {\it Upper-right:} $z=0$ and $R=20h^{-1}$Mpc; {\it Lower-left:}
  $z=1$ and $R=8h^{-1}$Mpc; {\it Lower-right:} $z=1$ and
  $R=20h^{-1}$Mpc.
\label{fig:prob_dh} }
\end{figure}

The ticks on the upper axis in each panel indicate the corresponding
conditional mean $\overline\deltah$ (eq.[\ref{eq:conditional_mean}]).
The peak position of the distribution is in reasonable agreement with
$\overline\deltah$.  As Figure \ref{fig:dhdm} indicates, $\halo$ given
$z$ and $R$ is fairly monotonically dependent on $M$ and $\zf$.  Thus
the peak in the conditional PDF reflects that of the formation epoch
distribution $\partial p/\partial \zf$. The width of the distribution
around the peak, on the other hand, is dominated by the mass
distribution since $\halo$ becomes more sensitive to the halo mass in
a denser environment (Fig.\ref{fig:dhdm}).

\subsection{Joint probability distribution of $\deltam$ and $\deltah$}
\label{subsec:jointPDF}

\begin{figure}
\begin{center}
   \leavevmode\epsfxsize=17cm \epsfbox{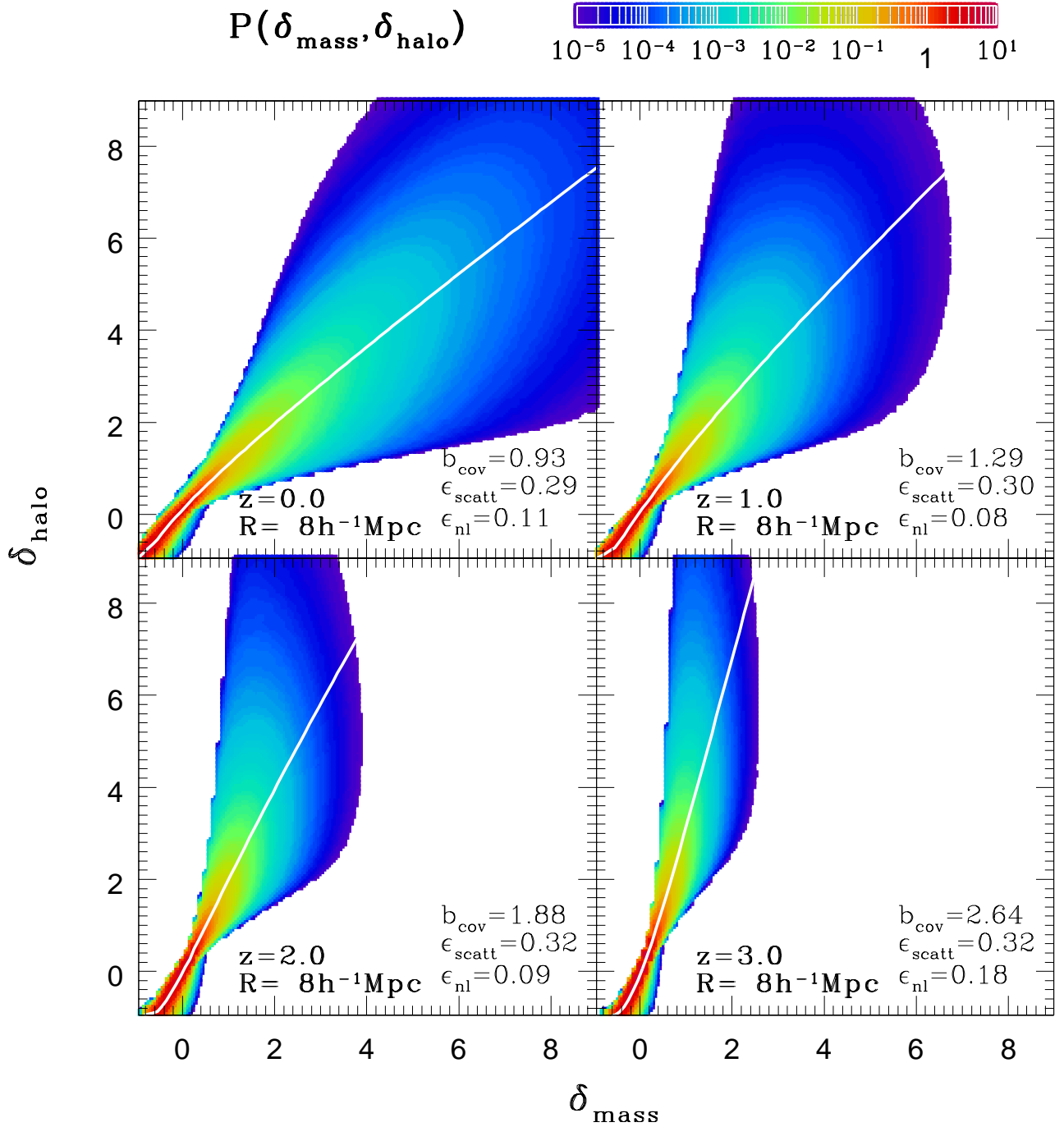}
\end{center}
\figcaption{Contour plot of the joint probability distribution of
  $\deltam$ and $\deltah$, $P(\deltam,\deltah)$. These examples are
  for $10^{11}h^{-1}M_\odot < M < 10^{13}h^{-1}M_\odot$ and
  $R=8h^{-1}$Mpc in LCDM model.  {\it Upper-left:} $z=0$; {\it
    Upper-right:} $z=1$; {\it Lower-left:} $z=2$; {\it Lower-right:}
  $z=3$.
\label{fig:dhdm_lincont} }
\end{figure}

Once $P(\deltah|\deltam)$ is given, the joint PDF $P(\deltam,\deltah)$
is simply obtained by multiplying the one-point PDF of the mass
density $P(\deltam)$, in our case, the log-normal model
(eq.[\ref{lognormalPDF}]). The resulting contour on
$(\deltam,\deltah)$ plane is illustrated in Figure
\ref{fig:dhdm_lincont}. This example shows the result with the top-hat
smoothing radius of $R=8h^{-1}$Mpc at different redshifts.  Solid
lines in each panel indicate the conditional mean $\overline\deltah
(\deltam)$.

The number density of halos of mass exceeding the current threshold
$M_{\rm min}=10^{11}h^{-1}\msun$ become progressively smaller as $z$
increases. Such halos naturally reside in higher density regions, and
therefore are strongly biased with respect to mass.  The biasing of those
halos gradually decreases as time since they simply follow the
gravitational field of the background mass after formation (Fry 1996;
Tegmark \& Peebles 1998; Taruya, Koyama \& Soda 1999). In addition,
new halos with $M>M_{\rm min}$ form more easily later and can be found
even at moderately dense environment.  For both reasons, the mean bias
$\overline\deltah$ as a function of $\deltam$ decreases at lower
redshifts.  

At $z=0$, the joint PDF $P(\deltam, \deltah)$ shows slightly 
anti-biasing behavior, i.e, $\overline\deltah\simlt\deltam$.  
This is partly due to the fact that a fraction of halos with $M<M_{\rm max}$ are
merged into a part of larger mass halos since the typical virialized
halo mass $M_*(z)$ at $z=0$ approaches the mass scale $M_0(R)$, 
corresponding to our adopted smoothing radius $R=8h^{-1}$Mpc itself.
In other words, our halo model generally predicts the positive-biasing
except for those halos of $M \simlt M_*(z) \simlt M_0(R)$ for given
$R$ and $z$.

\begin{figure}
\begin{center}
   \leavevmode\epsfxsize=17cm \epsfbox{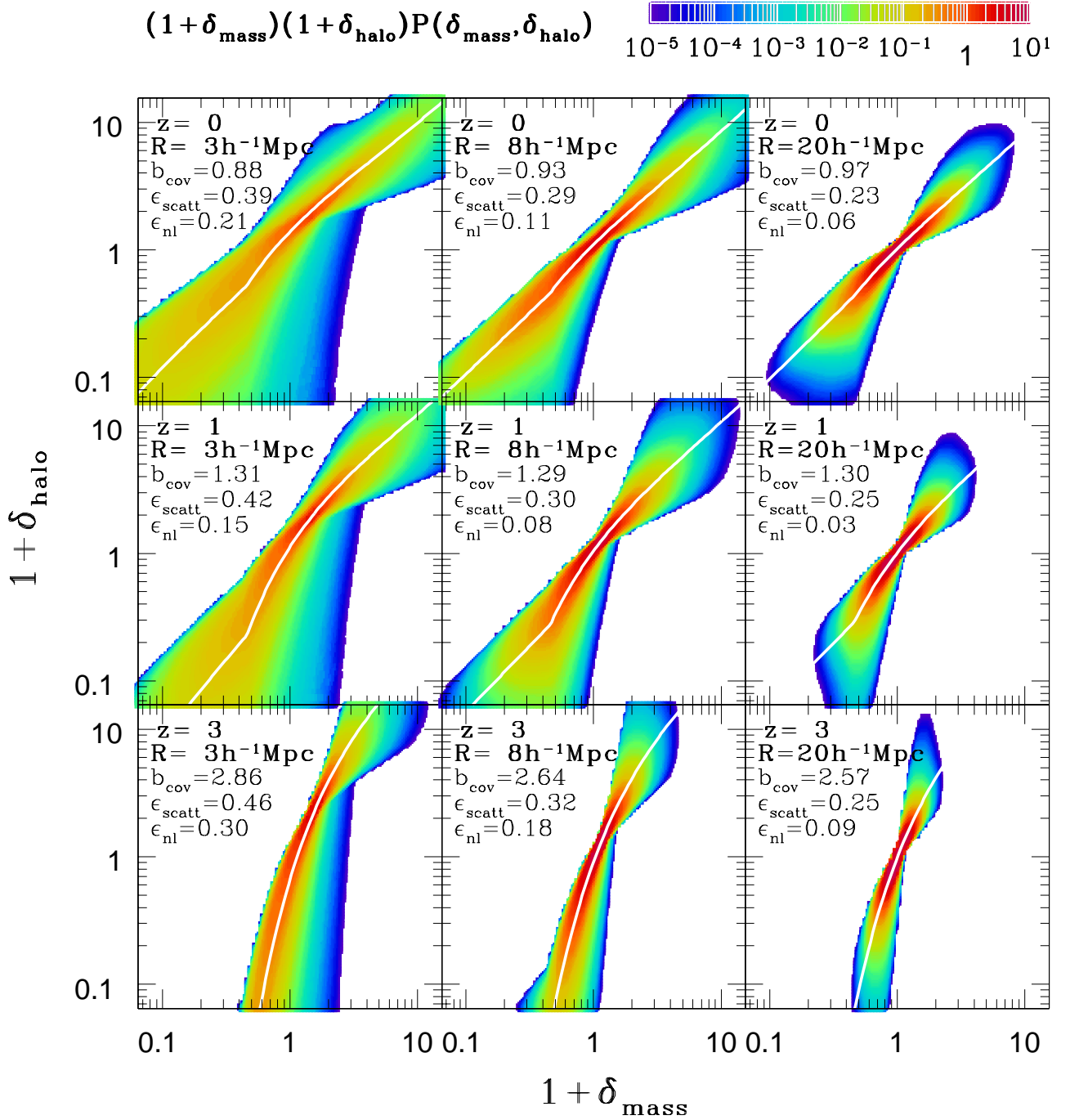}
\end{center}
\figcaption{Contour plot of the joint probability distribution of
  $\tilde P(1+\deltam,1+\deltah) =
  (1+\deltam)(1+\deltah)P(\deltam,\deltah)$. These examples are for
  $10^{11}h^{-1}M_\odot < M < 10^{13}h^{-1}M_\odot$ in LCDM model.
  The redshifts $z$ are $0$, $1$ and $3$ from top to bottom, and the
  smoothing lengths $R$ are $3h^{-1}$Mpc, $8h^{-1}$Mpc and
  $20h^{-1}$Mpc from left to right.
 \label{fig:dhdm_logcont} }
\end{figure}

While Figure \ref{fig:dhdm_lincont} elucidates the global feature of
the joint PDF, the statistical weights are practically dominated by
the relatively narrow regions around $\deltah \sim
\overline\deltah(\deltam)$. Those regions are illustrated better in
logarithmic scales rather than in linear scales. For this purpose, we
recompute the joint PDF from equations (\ref{weight}) and
(\ref{eq:constraint}) using the $3000\times3000$ mesh with the
logarithmically equal bin on the $(1+\deltag, 1+\deltam)$ plane.  Note
that the resulting PDF sampled in this way, $\tilde P(1+\deltam,
1+\deltah)$, satisfies
\begin{equation}
  \label{eq:linp2logp}
  \tilde P(1+\deltam, 1+\deltah) \, d \ln(1+\deltam) \, d \ln(1+\deltah)
= P(\deltam, \deltah) \, d \deltam \, d \deltah .
\end{equation}
Thus we decide to plot the joint PDF
$(1+\deltam)(1+\deltah)\,P(\deltam, \deltah)$. In Figure
\ref{fig:dhdm_logcont},  the dispersion around the mean biasing 
decreases at higher $z$ and/or larger $R$ in contrast to the
conditional PDF plotted in Figure \ref{fig:prob_dh}. This is basically
because the PDF $P(\deltam)$ in linear regime becomes
toward Gaussian and sharply peaked.

We remark that the contours of our joint PDF plotted in Figures
\ref{fig:dhdm_lincont} and \ref{fig:dhdm_logcont} seem to be very
narrow around $\deltam \sim \deltah \sim 0$.  This can be understood
by the fact that the positive halo density contrast preferentially
developed on the over-dense dark matter environment. In other words,
this is a natural consequence of our bias model in which the signs of
$\deltah$ and $\deltam$ are almost the same as illustrated in Figure
\ref{fig:dhdm}.  Incidentally this feature may be visually exaggerated
by the contours of very small probabilities; if one focuses only on
the contours of $P(\deltam,\deltah)>0.01$, the effect does not look so
strong.  In reality, and thus in numerical simulations, additional
stochastic processes other than the mass and formation epoch
distribution (including the dynamical motion of halos) should further
increase the scatter which makes the contour rounder than our
predictions.

We will return to these contour plots in understanding the behavior of
the biasing parameters later.

\subsection{Scale-dependence}
\label{subsec:scaledep}

The previous two subsections show that the joint and the conditional
PDFs are dependent on both $R$ and $z$. This dependence is translated
to the scale-dependence and redshift evolution of the second order
statistics defined in \S \ref{subsec:second-order}.

\begin{figure}
\begin{center}
   \leavevmode\epsfxsize=12cm \epsfbox{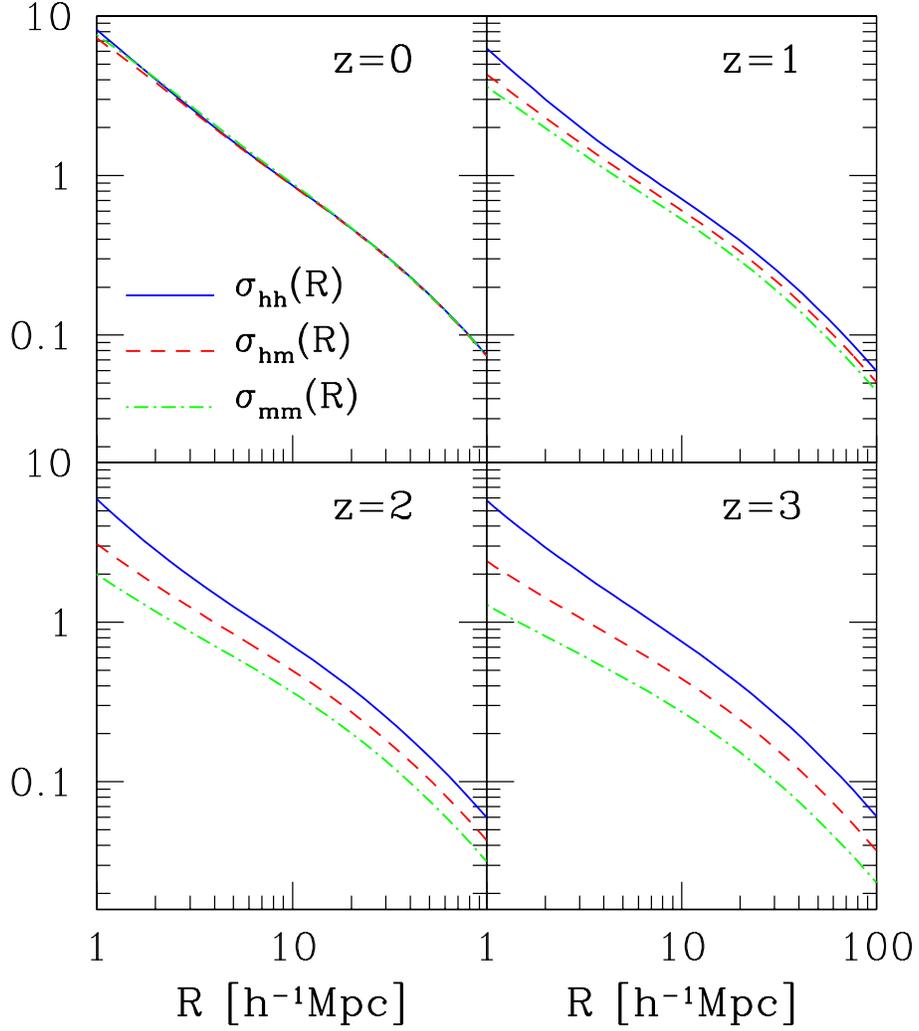}
\end{center}
\figcaption{Scale-dependence of $\sigma_{\rm hh}$ (solid), $\sigma_{\rm
hm}$ (dashed), and $\sigma_{\rm mm}$ (dot-dashed) at different
redshifts; {\it Upper-left:} $z=0$; {\it Upper-right:} $z=1$; {\it
Lower-left:} $z=2$; {\it Lower-right:} $z=3$.  \label{fig:sigma_hh} }
\end{figure}

\begin{figure}
\begin{center}
   \leavevmode\epsfxsize=12cm \epsfbox{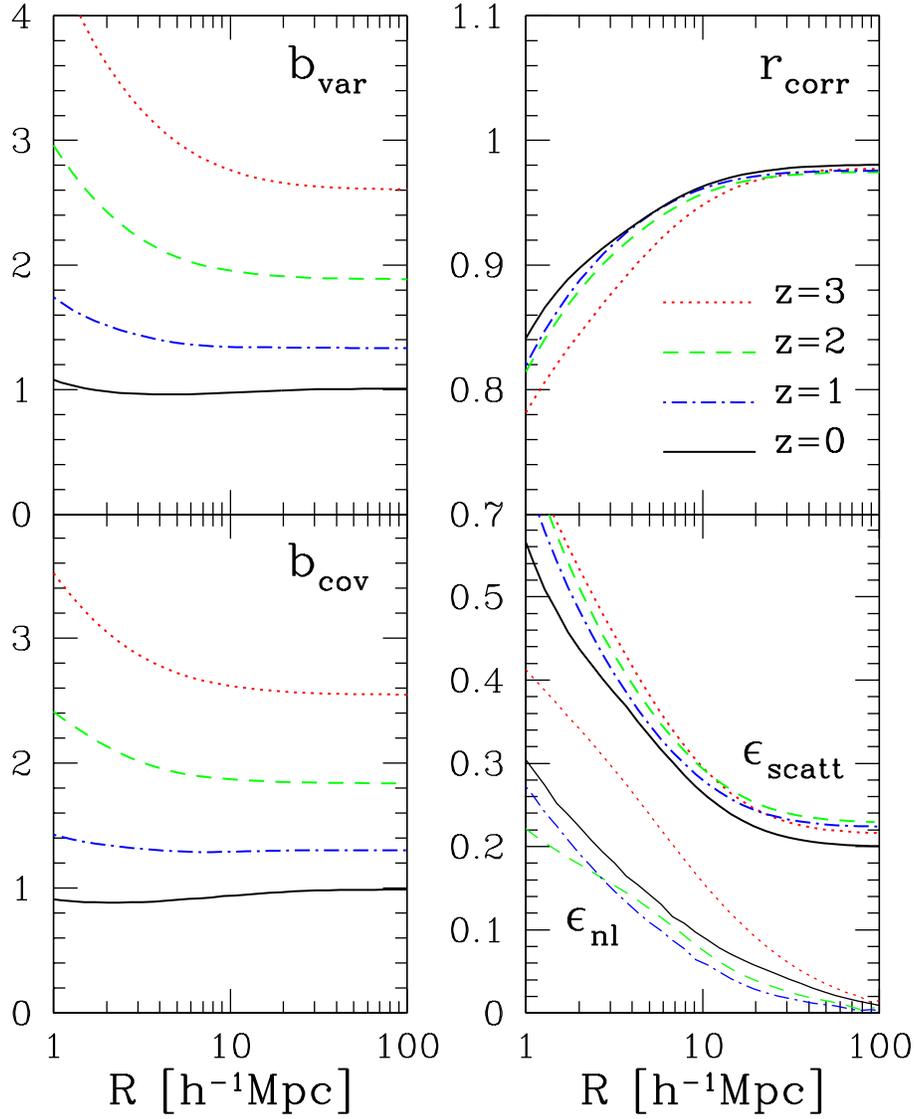}
\end{center}
\figcaption{Scale-dependence of biasing parameters, $\bvar$, $\rcorr$,
  $\bcov$, $\scatt$ and $\nonl$, at the different redshifts $z=0$
  (solid), $z=1$ (dot-dashed), $z=2$ (dashed), and $z=3$
  (dotted). In the {\it Lower-right} panel, thick and thin lines
  indicate the degree of stochasticity $\scatt$, and nonlinearity
  $\nonl$, respectively.  \label{fig:bias_r} }
\end{figure}

\begin{figure}
\begin{center}
   \leavevmode\epsfxsize=12cm \epsfbox{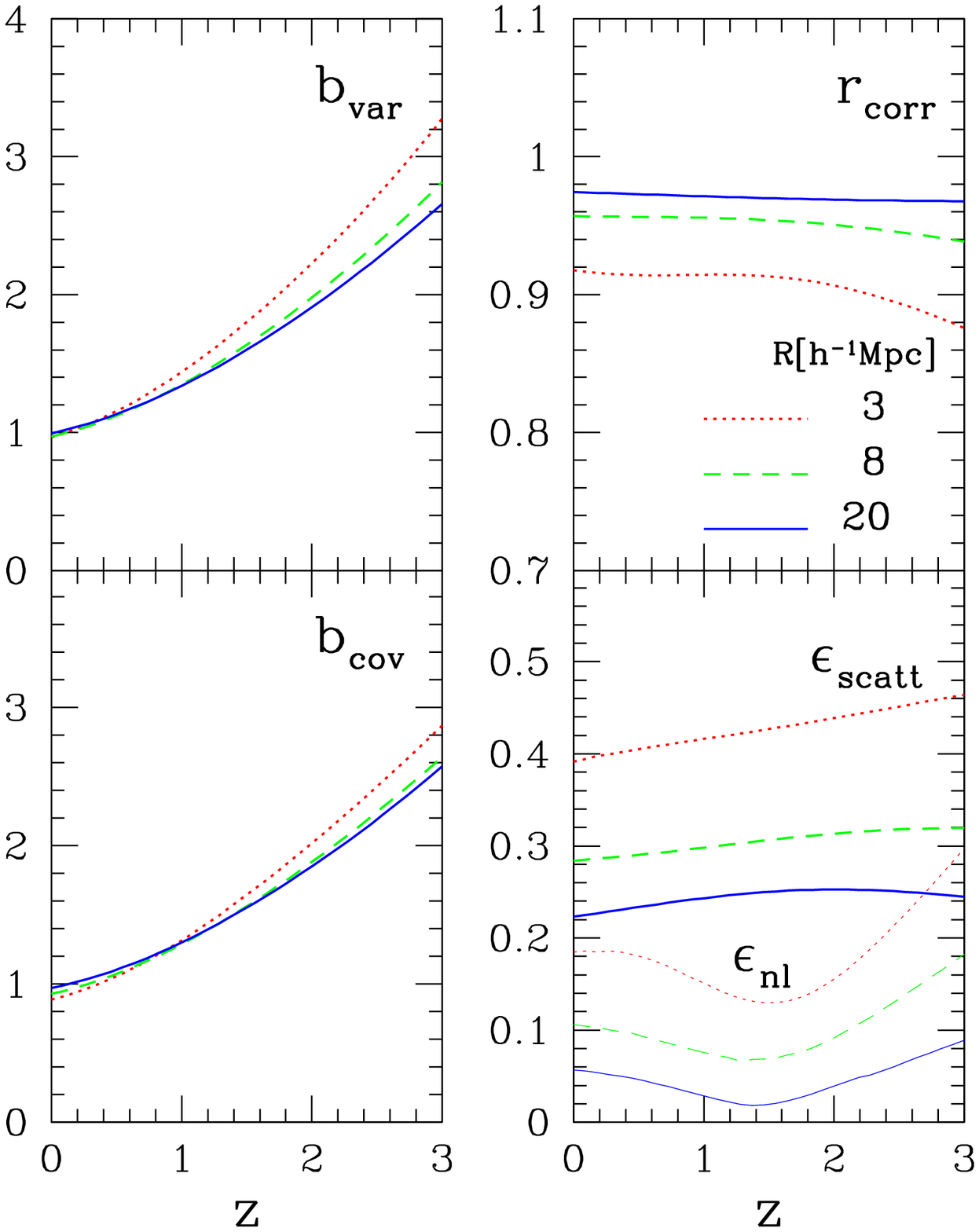}
\end{center}
\figcaption{Evolution of biasing parameters, $\bvar$, $\rcorr$, $\bcov$,
  $\scatt$ and $\nonl$, at $R=3\himpc$ (dotted), $8\himpc$
  (dashed), and $20\himpc$ (solid).  In the {\it
  Lower-right} panel, thick and thin lines indicate the degree of
  stochasticity $\scatt$, and nonlinearity $\nonl$, respectively.
  \label{fig:bias_z} }
\end{figure}

Figure \ref{fig:sigma_hh} shows $\sigmahh$, $\sigmahm$ and $\sigmamm$
at different redshifts as a function $R$. We compute those statistics
directly integrating equation (\ref{average}) over $M$, $\zf$ and
$\deltam$, instead of using $P(\deltam, \deltah)$.  While their scale-
and time-dependence is noticeable even from those panels, the biasing
parameters ($\bvar$, $\rcorr$, $\bcov$, $\scatt$ and $\nonl)$ plotted
in Figures \ref{fig:bias_r} and \ref{fig:bias_z} are more suitable in
understanding the origin of the behavior.

Consider first the scale-dependence (Fig.\ref{fig:bias_r}).  While
$\bvar$ and $\bcov$ are generally a decreasing function of $R$, this
behavior is significant only up to $R\simlt 2(1+z)\himpc$ in this
model.  This feature is more quantitatively exhibited by $\nonl$ and
$\scatt$ ({\it Lower-right} panel).  Therefore in practice the linear
biasing provides a good approximation on linear and quasi-linear
regimes. The biasing is non-deterministic especially on smaller scales
almost independently of $z$ (see also Fig.\ref{fig:bias_z}). In
addition, $\rcorr$ does not approach unity even on large scales,
implying that non-deterministic nature still exists there to some
extent. As the lower-right panel in Figure \ref{fig:bias_r} indicates,
$\scatt \gg \nonl$ on all scales and the above feature should be
ascribed to the stochasticity due to the distribution of $M$ and
$\zf$. In fact, this stochasticity on large scales is expressed
explicitly in terms of the linear biasing approximation by Mo \& White
(1996):
\begin{eqnarray}
\label{eq:MWbias}
\deltah \simeq b_{\MW}~\deltam , \qquad
b_{\MW}(z|M,\zf) = 1+\frac{\nu^{2}-1}{\delta_{c}(z,\zf)} , 
\end{eqnarray}
with $\nu \equiv \delta_{c}(z,\zf)/\sigma(M,z)$.  It should be
noted that $b_{\MW}$ is often regarded as a function of $z$ and $M$
assuming $z=\zf$, leading to the linear {\it deterministic} model. Thus
once a halo mass $M$ is specified, $\scatt=0$. In equation
(\ref{eq:MWbias}), however, we explicitly keep the $\zf$-dependence
which adds the stochastic nature in the model. More specifically, the
definition (\ref{eq:scatt}) with equations (\ref{average}) and
(\ref{eq:MWbias}) reduces to
\begin{equation}
 \label{eq:scatt_eff}
\scatt^{2} 
= \frac{\llangle\deltam^{2}\rrangle \,
      \llangle(b_{\MW}\deltam- 
  \llangle b_{\MW} \rrangle_{M,\zf} \deltam)^2\rrangle}
    {\llangle \llangle b_{\MW} \rrangle_{M,\zf} \deltam^2\rrangle^2} 
= \frac{\llangle b_{\MW}^{2}\rrangle_{M,\zf}}
     {\llangle b_{\MW} \rrangle_{M,\zf}^{2}} - 1 ,
\end{equation}
where $\llangle \cdots \rrangle_{M,\zf}$ denotes the average over $M$
and $\zf$. This accounts for the scale-independent non-vanishing
$\scatt$ exhibits in Figure \ref{fig:bias_r}.  In conclusion, our
model implies that the halo biasing does {\it not} become fully
deterministic even on large scales where its nonlinearity is
negligible.  This result is not surprising since we take into account
the stochastic processes which do not vanish on large scales, but the
overall effect is quite small ($\rcorr \sim 0.98$).

\subsection{Redshift dependence}
\label{subsec:zdep}

Next discuss the redshift dependence of our biasing model.  Figure
\ref{fig:bias_z} shows that $\bvar$ and $\bcov$ strongly evolve in
time.  In fact, this is in marked contrast with predictions on the
basis of phenomenological {\it linear deterministic} models. For
example, a model of Fry (1996) leads to evolution of a form:
\begin{equation} 
\label{eq:linearbz}
  b(z)= 1 +{1\over D(z)} [b(z=0)-1] .
\end{equation}
This implicitly assumes that all the objects of interest form at the
same $\zf$ and that their biasing parameters at $\zf$ are independent
of the mass. Since neither the above assumptions apply to our model,
the prediction (\ref{eq:linearbz}) is quite different from ours even
in the linear regime. Our results generally show much stronger
evolution as $z$ increases despite the fact that $b(z=0)$ is very
close to unity. The recent compilation of the various galaxy catalogs
also indicates that the prediction (\ref{eq:linearbz}) does not
reasonably describe the behavior at $z\simgt2$ (Magliocchetti et al.
1999). Thus, the proper modeling in the framework of the nonlinear
stochastic biasing is important even in predicting $\bvar$ and
$\bcov$.

In our halo biasing model, the degree of stochasticity $\scatt$ is
almost constant in time because it is determined by the effective
widths of the probability distribution functions of $M$ and $\zf$.
Incidentally the nonlinearity $\nonl$ does not evolve monotonically
(thin lines in the {\it Lower-right} panel). At an intermediate
redshift, $\nonl$ reaches at a minimum. This behavior is qualitatively
explained from the curvature of the conditional mean
$\overline\deltah$ as a function of $\deltam$, i.e., its second
derivative $d^2 \overline\deltah/d\deltam^2$.  At $\deltam \gg 1$,
halos of the mass $M_{\rm min} < M < M_{\rm max}$ that we adopt here
exhibit stronger positive biasing ($\overline\deltah > \deltam$) on
average at $z\simgt 2$ and mildly anti-biasing ($\overline\deltah
\simlt \deltam$) at $z\simlt 1$. Since $\deltah=-1$ at $\deltam=-1$ by
definition (cf., eq.[\ref{eq:delta-halo}]), the dependence results in
positive and negative curvatures of $\overline\deltah(\deltam)$,
respectively at $z\simgt 2$ and $z\simlt 1$, especially around
$\deltam \sim 0$ where the contribution of the joint PDF
$P(\deltam,\deltah)$ is significant. This feature is clearly visible
in Figure \ref{fig:dhdm_lincont}. In turn, the curvature should be
minimum somewhere between $z=1$ and $2$. When higher-order correction
is neglected, $\nonl$ is dominated by the curvature or the second
derivative of $\overline\deltah(\deltam)$ properly averaged over
$\deltam$ (eq.[\ref{eq:nonl}]), and it should become minimum at the
same redshift. It is interesting to note that the qualitatively
similar evolutionary feature was found from the numerical simulations
of Somerville et al. (2000; their Fig.17).

\subsection{Origin of stochasticity and cosmological model dependence}
\label{subsec:modeldep}

\begin{figure}
\begin{center}
   \leavevmode\epsfxsize=16.5cm \epsfbox{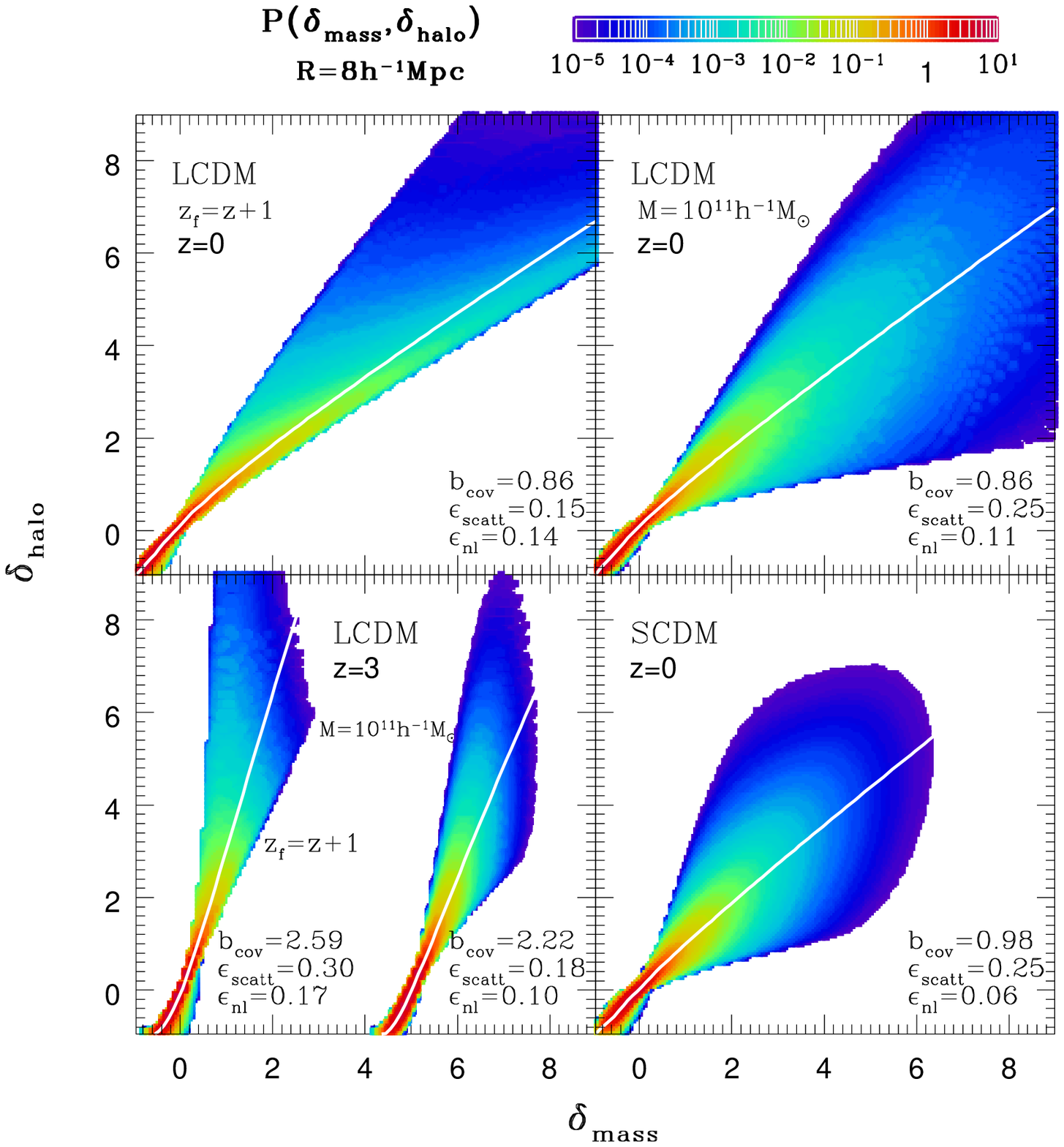}
\end{center}
\figcaption{Joint probability distribution of $\deltam$ and $\deltah$,
  $P(\deltam,\deltah)$ at $R=8h^{-1}$Mpc for different models. {\it
    Upper-left:} LCDM with $ z_{\rm f} = z + 1$ at $z=0$; {\it
    Upper-right:} LCDM with $M = 10^{11}h^{-1}\msun$ at $z=0$; {\it
    Lower-left:} LCDM with $ z_{\rm f} = z + 1$ and with
  $M=10^{11}h^{-1}M_\odot$ (the value of $\deltam$ is artificially
  shifted by five for clarity), at $z=3$;  {\it Lower-right:} SCDM
  with $10^{11}h^{-1}\msun < M < 10^{13}h^{-1}\msun$ at $z=0$.
 \label{fig:dhdm_lincont_diff} }
\end{figure}

\begin{figure}
\begin{center}
   \leavevmode\epsfxsize=12cm \epsfbox{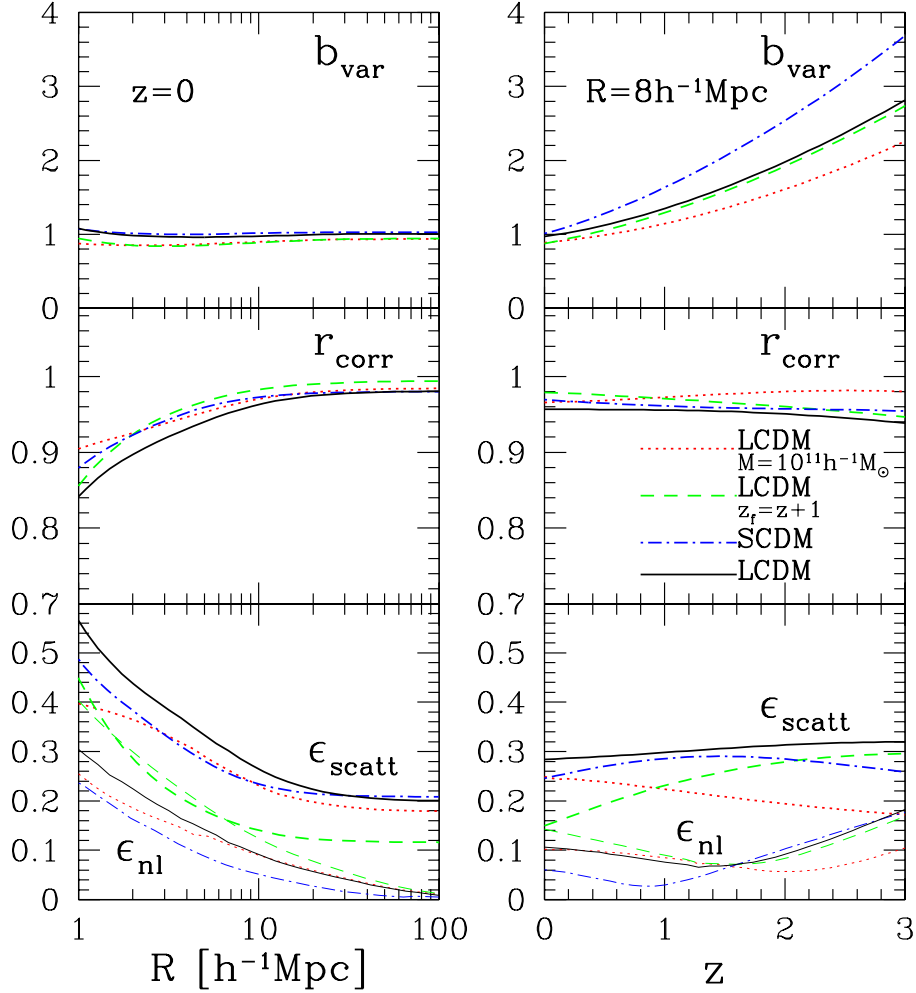}
\end{center}
\figcaption{Comparison of $\bvar$, $\rcorr$, $\scatt$ and $\nonl$ in
  different models. {\it Left:} at $z=0$; {\it Right:} at $R=8\himpc$.
  Different lines indicate results for LCDM with $ M =
  10^{11}h^{-1}M_\odot$ (dotted), LCDM with $ z_{\rm f} = z + 1$
  (dashed), and SCDM with $10^{11}h^{-1}M_\odot < M <
  10^{13}h^{-1}M_\odot$ (dot-dashed), in comparison with our
  canonical model (LCDM with $10^{11}h^{-1}M_\odot < M <
  10^{13}h^{-1}M_\odot$) plotted in solid lines.
  \label{fig:bias_diff} }
\end{figure}

So far we have presented model predictions for halos averaged over
$10^{11}h^{-1}M_\odot < M < M^{13}h^{-1}M_\odot$ in LCDM model taking
into account the appropriate $\zf$ distribution. Figures
\ref{fig:dhdm_lincont_diff} and \ref{fig:bias_diff} compare those
fiducial results with predictions based on the different model
assumptions.

First we address the question of the origin of the stochasticity.
Since our biasing model has two {\it hidden} parameters, $M$ and
$\zf$, we attempt to separate the two sources by fixing
$M=10^{11}h^{-1}M_\odot$ or $\zf=z+1$ while keeping the other
parameters exactly the same. The upper-panels and the lower-left panel
of Figure \ref{fig:dhdm_lincont_diff} suggest that the $\zf$
distribution dominates the stochasticity at low redshift $(z=0)$,
while the effect of the $M$ distribution becomes significant at higher
redshift $(z=3)$. The same behaviors can be seen in the lower-right
panel of Figure \ref{fig:bias_diff}. Since our model relies on the
hierarchical picture of structure formation, the result is simply
deduced from the merging history of halos.  Thus, in general, the
major contribution to the joint PDF can become the formation epoch
distribution. This is also indicated in the scale-dependence of the
stochasticity in the lower-left panel of Figure
\ref{fig:bias_diff}(thick-dashed and thick-dotted lines).

Next consider the cosmological model dependence.  For this purpose, we
plot the result in the SCDM model with the same mass range.  The joint
PDF at $z=0$ and $R=8\himpc$ ({\it Lower-right} panel of
Fig.\ref{fig:dhdm_lincont_diff}) is confined in a slightly narrower
regime compared with that in LCDM. This comes from the fact that the
formation epoch in SCDM shows a bit more sharply peaked distribution
$\partial p/\partial \zf$ in than that in LCDM with the same halo mass
$M$ (Fig.\ref{fig:dpdzf}).  As a result, the stochasticity in SCDM is
smaller (i.e., $\scatt$ is smaller and $\rcorr$ is closer to unity), but
$\bvar$ at $z=0$ is almost insensitive to such small changes. Rather the
major difference between LCDM and SCDM is the redshift evolution of
$\bvar$ which increases more rapidly in SCDM reflecting the faster
growth rate of density fluctuations.

\subsection{Comparison with previous results}
\label{subsec:comparison}

Dark matter halos are quite natural and likely sites for galaxy and
cluster formation. Thus there are many previous papers to discuss
different aspects of the halo biasing on the basis of different
assumptions and modeling (Catelan et al. 1999a,b; Blanton et al.
1999). Among others, Kravtsov \& Klypin (1999) and Somerville et al.
(2000) analyzed the nonlinearity and stochasticity in halo and galaxy
biasing using numerical simulations.  In this sense their work is
complementary to our analytical modeling, and deserves quantitative
comparison with our results.

Kravtsov \& Klypin (1999) performed high-resolution N-body simulations
employing $N=256^3$ particles in a periodic $60\himpc$ box so as to
overcome the halo over-merging. In particular, their Figure 4 plotting
the joint PDF $P(\deltam,\deltah)$ is quite relevant for the
comparison with our Figure \ref{fig:dhdm_lincont}. Strictly speaking,
their simulated halo catalogue is based on slightly different
identification scheme (Klypin et al. 1999); the {\it bound density
  maxima} algorithm, a selected mass range $[M_{\rm min}, M_{\rm
  max}]$ is limited both by the numerical resolution (individual
particle mass) and the simulation boxsize. Furthermore their
statistics is based on the smoothing length $R=5h^{-1}$Mpc.  Despite
such quantitative difference, both contours look surprisingly similar.
They pointed out that the Mo \& White biasing with $\zf=z+1$
phenomenologically fits the conditional mean
$\overline\deltah(\deltam)$ of their simulations. Actually this is
exactly what we find here, and mainly due to the fairly strong peak in
the formation epoch distribution (Fig. \ref{fig:dpdzf}).

Somerville et al. (2000) examined the nonlinear and stochastic nature
in the galaxy biasing combining N-body simulations and {\it
  semi-analytic} model of galaxy formation (see also Kauffmann, Nusser
\& Steinmetz 1997). While their model incorporates many detailed
astrophysical processes (gas cooling, star formation, stellar
evolution and population synthesis, galaxy feedback and merging among
others) that our analytical modeling neglects, their overall
conclusion is that the massive halos ($M\simgt 10^{11}h^{-1} M_\odot$)
identified from simulations have nearly one-to-one correspondence with
luminous {\it galaxies}.  Their conclusion is encouraging to us since
it justifies our crucial assumption that the nonlinearity and
stochasticity in the halo biasing are originated mainly from
gravitational clustering. In fact, their Figs. 5 and 6 are also very
similar to our joint PDF plotted in Figure \ref{fig:dhdm_logcont}.
Furthermore they find (their Fig.17) that both nonlinearity and
stochasticity of galaxies evolve moderately as redshift, and decrease
on larger scales, and that the stochasticity shows a minimum at an
intermediate redshift.  These findings are fully consistent with our
results and in fact can be explained physically in the framework of
our analytic description.

\section{Discussion and Conclusions}
\label{sec:conclusions}

In this paper, we presented a general formalism to describe the
nonlinear stochastic biasing from the hidden variable interpretation.
According to this formulation we proposed a physical model for the
halo biasing assuming that mass and formation epoch of dark halos act
as the major hidden variables. In particular, this model for the first
time provides an analytic expression for the joint probability
distribution function of the halos and the dark matter density fields.
The specific functional forms for the PDFs can be derived, or are at
least consistent with the phenomenological results, from the standard
gravitational instability theory and the assumption of the
random-Gaussianity of the primordial density field.  Therefore we
expect that the basic features of the nonlinear and stochastic biasing
predicted from our model are fairly generic. In fact, detailed
comparison with the previous numerical simulations by Kravtsov \&
Klypin (1999) and Somerville et al. (2000) shows good agreement with
our predictions, indicating that the nonlinear and stochastic nature
of the halo biasing is essentially understood by taking account of the
distribution of the halo mass and the formation epoch.

Then we introduced a set of biasing parameters from second-order
statistical quantities following Dekel \& Lahav (1999), which properly
quantify the one-point statistics of nonlinear and stochastic biasing.
As specific examples, we explicit compute those parameters in LCDM
model, and examined their scale- and time-dependence. Our major findings
are summarized as follows:
\begin{enumerate}
\item The biasing of the variance, $\bvar \equiv \sqrt{\sigmahh/\sigmamm}$,
      evolves strongly as redshift. While its scale-dependence becomes
      noticeable on small scales and/or at high redshifts, a simple
      linear biasing model provides a reasonable approximation
      roughly at $R\simgt 2(1+z)\himpc$.
\item The stochasticity, $\rcorr \equiv
      \sigmahm^2/\sqrt{\sigmahh\sigmamm}$ exhibits moderate
      scale-dependence especially on $R\simlt 20\himpc$, but is almost
      independent of $z$. Since $\scatt \gg \nonl$ in general, the
      stochasticity of halo biasing is mainly generated by the scatter
      which is dominated by the formation epoch distribution at lower
      redshifts and by the halo mass distribution at
      higher redshifts. Importantly, the stochasticity 
      still remains constant even on large 
      scales, which yields the small deviation of $\rcorr$ from unity. 
\end{enumerate}
The fact that biasing approaches linear and scale-independent on large
scales is just expected from previous theoretical work (e.g., Coles 1993;
Scherrer \& Weinberg 1998) and was recently suggested observationally
(Seaborne et al. 1999). Our model further implies that the
scale-independence still holds even in the quasi-linear regime, and
more importantly, that biasing is still stochastic to some extent on
those scales.

Before closing we would like to comment on several limitations of our
model and on future prospects in order. First, our analysis is
entirely dependent on the formation redshift distribution derived by
Lacey \& Cole (1993). It is known, however, that the distribution
becomes negative in some models, suggesting that this function is not
fully correct. Another proposal for the distribution function by
Kitayama \& Suto (1996a) also suffers from  a different conceptual
problem. This might be ascribed to the difficulty of distinguishing
the merger and accretion unambiguously within the framework of the
extended PS theory.  Despite this limitation, the two proposals by
Lacey \& Cole (1993) and by Kitayama \& Suto (1996a) lead to very
similar predictions and thus we believe it unlikely that this
significantly affects the final results.  Second, several authors have
claimed that a better agreement with numerical simulations is attained
by incorporating the non-spherical effect into the Press-Schechter
theory (Lee \& Shandarin 1998; Sheth, Mo, \& Tormen 1999). Although we
do not think that the non-spherical effect drastically changes the
statistical predictions of our model, this is definitely a
well-defined and interesting possibility to improve our model provided
that the corresponding distribution function of the formation epoch
can be derived. Third, for the proper comparison with observation, our
predictions should be corrected for the redshift-space distortion.
Since the distortion induced by peculiar velocity field also 
leads to the stochasticity, the resulting biasing becomes 
object-dependent (Taruya et al. 2000).  Finally and
more importantly, the astrophysical effects other than gravity must be
taken into account. Since a halo identified in our scheme carries its
formation epoch, it is not so difficult to combine with the
phenomenological galaxy formation model.  Such an improved model will
enable more generic and testable predictions for the {\it galaxy}
biasing.

\bigskip
\bigskip

A.T. gratefully acknowledges support from a JSPS (Japan Society for
the Promotion of Science) fellowship.  Numerical computations were
carried out at RESCEU (Research Center for the Early Universe,
University of Tokyo) and ADAC (the Astronomical Data Analysis Center)
of the National Astronomical Observatory, Japan.  This research was
supported in part by the Grant-in-Aid by the Ministry of Education,
Science, Sports and Culture of Japan (07CE2002) to RESCEU.

\bigskip \bigskip



\clearpage

\centerline{\bf Appendices}

\appendix

\section{Mass variance for the CDM model}
\label{appendix:sigmam}

The mass variance $\sigma(M,z=0)$ defined by equation
(\ref{mass-variance}) requires the linear power spectrum of density
fluctuations $P_{\rm lin}(k)$. The fitting form of the CDM power
spectrum is given by Bardeen et al. (1986) with the scale-invariant
Harrison-Zel'dovich initial condition:
\begin{equation}
  P_{\rm lin}(k) \propto k \left[\ln(1+2.34q) \over 2.34 q \right]^2 
  \left[1 + 3.89 q +
    (16.1q)^2 + (5.46q)^3 + (6.71q)^4\right]^{-1/2},
\label{eq:cdmspect}
\end{equation}
where $q\equiv k/(\Gamma h \mbox{ Mpc}^{-1})$. Here, we adopt
$\Gamma=\Omega_0 h \exp(-\Omega_{\rm b}-\sqrt{2h}\Omega_{\rm
  b}/\Omega_0)$ with the baryon density parameter $\Omega_{\rm
  b}=0.02h^{-2}$.

Numerical integration of equation (\ref{mass-variance}) is
straightforward, but time-consuming since we heavily use $\sigma^2$ as
well as its derivative $d\sigma^2/dM$. Fortunately Kitayama \& Suto
(1996b) obtained the following accurate fitting formula for $\sigma^2$
whose derivative simultaneously fits $d\sigma^2/dM$:
\begin{equation}
  \sigma \propto \left(1 + 2.208 m^p - 0.7668 m^{2p} +
    0.7949 m^{3p}\right)^{-2 /(9p)},
\label{eq:cdmfit}
\end{equation}
where $p=0.0873$, and $m\equiv M(\Gamma h)^2/(10^{12}\msun)$.  
The above approximation holds within a 
few percent for both $\sigma^2$ and $d\sigma^2/dM$ in the range 
$10^{-6} \simlt m \simlt 10^{4}$.

The normalization of $\sigma^2$ is characterized by the parameter
$\sigma_{8}$:
\begin{equation}
\sigma(R_{\rm M}= 8h^{-1}\mbox{Mpc},z=0) = \sigma_{8},  
\label{eq:bias}
\end{equation}
where $R_{\rm M}=[3M/(4\pi \bar{\rho}_{\rm mass})]^{1/3}$.  Throughout
the paper we adopt the above formula (\ref{eq:cdmfit}) combined with the
cluster normalization for $\sigma_8$ (Kitayama \& Suto 1997).

\section{Fitting formulae for the distribution function of 
  halo formation epoch}
\label{appendix:zf}

The distribution function of the halo formation epoch
(eq.[\ref{dp-domegaf}]) plays a central role in our model, but it
requires a time-consuming numerical integration and inversion.  Thus in
the present paper we use the following fitting formulae of Kitayama \&
Suto (1996b):
\begin{eqnarray}
\label{eq:dpdwfit}
{\partial p \over \partial \tilde \omega_{\rm f}} (\alpha,
\tilde\omega_{\rm f}) &=& { A(\alpha) \over 1 + B(\alpha)\tilde
  \omega_{\rm f}} {\rm e}^{-5\tilde \omega_{\rm f}^2} + 2 C(\alpha)
\tilde \omega_{\rm f} \, {\rm erfc}\left( {\tilde \omega_{\rm f} \over
    \sqrt{2}} \right), 
\end{eqnarray}
where ${\rm erfc}(x)$ is the complementary error function, and
\begin{eqnarray}
  A(\alpha) &\equiv& \sqrt{8 \over \pi}(1-\alpha)(0.0107 +
  0.0163\alpha), \\ 
  B(\alpha) &\equiv& {2 \over A(\alpha)} \left[ C(\alpha) -
    {2^\alpha-1 \over \alpha} \right] , \\ 
  C(\alpha) &\equiv& 1 - {1-\alpha \over 25}.
\end{eqnarray}
The parameter $\alpha$ is related to the spectral index of the mass
variance $\sigma(M)$. Kitayama \& Suto (1996b) showed that in the CDM
model this parameter should be replaced by
\begin{equation}
\label{eq:alphaeff}
\alpha = \alpha_{\rm eff}(0.6268 + 0.3058
\alpha_{\rm eff}) ,
\end{equation}
where the effective spectral index $\alpha_{\rm eff} \equiv -
d\log\sigma_{\rm CDM}^2/d\log M$ is computed from the fitting formula
(\ref{eq:cdmfit}) and its derivative.

\clearpage

\clearpage

\clearpage

\clearpage

\clearpage

\end{document}